
\catcode `\@=11 

\def\@version{1.3}
\def\@verdate{28.11.1992}


%
%
%
%
%
%

\font\fiverm=cmr5
\font\fivei=cmmi5	\skewchar\fivei='177
\font\fivesy=cmsy5	\skewchar\fivesy='60
\font\fivebf=cmbx5

\font\sevenrm=cmr7
\font\seveni=cmmi7	\skewchar\seveni='177
\font\sevensy=cmsy7	\skewchar\sevensy='60
\font\sevenbf=cmbx7

\font\eightrm=cmr8
\font\eightbf=cmbx8
\font\eightit=cmti8
\font\eighti=cmmi8			\skewchar\eighti='177
\font\eightmib=cmmib10 at 8pt	\skewchar\eightmib='177
\font\eightsy=cmsy8			\skewchar\eightsy='60
\font\eightsyb=cmbsy10 at 8pt	\skewchar\eightsyb='60
\font\eightsl=cmsl8
\font\eighttt=cmtt8			\hyphenchar\eighttt=-1
\font\eightcsc=cmcsc10 at 8pt
\font\eightsf=cmss8

\font\ninerm=cmr9
\font\ninebf=cmbx9
\font\nineit=cmti9
\font\ninei=cmmi9			\skewchar\ninei='177
\font\ninemib=cmmib10 at 9pt	\skewchar\ninemib='177
\font\ninesy=cmsy9			\skewchar\ninesy='60
\font\ninesyb=cmbsy10 at 9pt	\skewchar\ninesyb='60
\font\ninesl=cmsl9
\font\ninett=cmtt9			\hyphenchar\ninett=-1
\font\ninecsc=cmcsc10 at 9pt
\font\ninesf=cmss9

\font\tenrm=cmr10
\font\tenbf=cmbx10
\font\tenit=cmti10
\font\teni=cmmi10		\skewchar\teni='177
\font\tenmib=cmmib10	\skewchar\tenmib='177
\font\tensy=cmsy10		\skewchar\tensy='60
\font\tensyb=cmbsy10	\skewchar\tensyb='60
\font\tenex=cmex10
\font\tensl=cmsl10
\font\tentt=cmtt10		\hyphenchar\tentt=-1
\font\tencsc=cmcsc10
\font\tensf=cmss10

\font\elevenrm=cmr10 scaled \magstephalf
\font\elevenbf=cmbx10 scaled \magstephalf
\font\elevenit=cmti10 scaled \magstephalf
\font\eleveni=cmmi10 scaled \magstephalf	\skewchar\eleveni='177
\font\elevenmib=cmmib10 scaled \magstephalf	\skewchar\elevenmib='177
\font\elevensy=cmsy10 scaled \magstephalf	\skewchar\elevensy='60
\font\elevensyb=cmbsy10 scaled \magstephalf	\skewchar\elevensyb='60
\font\elevensl=cmsl10 scaled \magstephalf
\font\eleventt=cmtt10 scaled \magstephalf	\hyphenchar\eleventt=-1
\font\elevencsc=cmcsc10 scaled \magstephalf
\font\elevensf=cmss10 scaled \magstephalf

\font\fourteenrm=cmr10 scaled \magstep2
\font\fourteenbf=cmbx10 scaled \magstep2
\font\fourteenit=cmti10 scaled \magstep2
\font\fourteeni=cmmi10 scaled \magstep2		\skewchar\fourteeni='177
\font\fourteenmib=cmmib10 scaled \magstep2	\skewchar\fourteenmib='177
\font\fourteensy=cmsy10 scaled \magstep2	\skewchar\fourteensy='60
\font\fourteensyb=cmbsy10 scaled \magstep2	\skewchar\fourteensyb='60
\font\fourteensl=cmsl10 scaled \magstep2
\font\fourteentt=cmtt10 scaled \magstep2	\hyphenchar\fourteentt=-1
\font\fourteencsc=cmcsc10 scaled \magstep2
\font\fourteensf=cmss10 scaled \magstep2

\font\seventeenrm=cmr10 scaled \magstep3
\font\seventeenbf=cmbx10 scaled \magstep3
\font\seventeenit=cmti10 scaled \magstep3
\font\seventeeni=cmmi10 scaled \magstep3	\skewchar\seventeeni='177
\font\seventeenmib=cmmib10 scaled \magstep3	\skewchar\seventeenmib='177
\font\seventeensy=cmsy10 scaled \magstep3	\skewchar\seventeensy='60
\font\seventeensyb=cmbsy10 scaled \magstep3	\skewchar\seventeensyb='60
\font\seventeensl=cmsl10 scaled \magstep3
\font\seventeentt=cmtt10 scaled \magstep3	\hyphenchar\seventeentt=-1
\font\seventeencsc=cmcsc10 scaled \magstep3
\font\seventeensf=cmss10 scaled \magstep3

\def\@typeface{Computer Modern} 

\def\hexnumber@#1{\ifnum#1<10 \number#1\else
 \ifnum#1=10 A\else\ifnum#1=11 B\else\ifnum#1=12 C\else
 \ifnum#1=13 D\else\ifnum#1=14 E\else\ifnum#1=15 F\fi\fi\fi\fi\fi\fi\fi}

\def\mib{\hexnumber@\mibfam}
\def\syb{\hexnumber@\sybfam}

\def\makestrut{%
  \setbox\strutbox=\hbox{%
    \vrule height.7\baselineskip depth.3\baselineskip width 0pt}%
}

\def\bls#1{%
  \normalbaselineskip=#1%
  \normalbaselines%
  \makestrut%
}

%

\newfam\mibfam 
\newfam\sybfam 
\newfam\scfam  
\newfam\sffam  

\def\em{\ifdim\fontdimen1\font>0 \rm\else\it\fi}

\textfont3=\tenex
\scriptfont3=\tenex
\scriptscriptfont3=\tenex

\def\eightpoint{
  \def\rm{\fam0\eightrm}%
  \textfont0=\eightrm \scriptfont0=\sevenrm \scriptscriptfont0=\fiverm%
  \textfont1=\eighti  \scriptfont1=\seveni  \scriptscriptfont1=\fivei%
  \textfont2=\eightsy \scriptfont2=\sevensy \scriptscriptfont2=\fivesy%
  \textfont\itfam=\eightit\def\it{\fam\itfam\eightit}%
  \textfont\bffam=\eightbf%
    \scriptfont\bffam=\sevenbf%
      \scriptscriptfont\bffam=\fivebf%
  \def\bf{\fam\bffam\eightbf}%
  \textfont\slfam=\eightsl\def\sl{\fam\slfam\eightsl}%
  \textfont\ttfam=\eighttt\def\tt{\fam\ttfam\eighttt}%
  \textfont\scfam=\eightcsc\def\sc{\fam\scfam\eightcsc}%
  \textfont\sffam=\eightsf\def\sf{\fam\sffam\eightsf}%
  \textfont\mibfam=\eightmib%
  \textfont\sybfam=\eightsyb%
  \bls{10pt}%
}

\def\ninepoint{
  \def\rm{\fam0\ninerm}%
  \textfont0=\ninerm \scriptfont0=\sevenrm \scriptscriptfont0=\fiverm%
  \textfont1=\ninei  \scriptfont1=\seveni  \scriptscriptfont1=\fivei%
  \textfont2=\ninesy \scriptfont2=\sevensy \scriptscriptfont2=\fivesy%
  \textfont\itfam=\nineit\def\it{\fam\itfam\nineit}%
  \textfont\bffam=\ninebf%
    \scriptfont\bffam=\sevenbf%
      \scriptscriptfont\bffam=\fivebf%
  \def\bf{\fam\bffam\ninebf}%
  \textfont\slfam=\ninesl\def\sl{\fam\slfam\ninesl}%
  \textfont\ttfam=\ninett\def\tt{\fam\ttfam\ninett}%
  \textfont\scfam=\ninecsc\def\sc{\fam\scfam\ninecsc}%
  \textfont\sffam=\ninesf\def\sf{\fam\sffam\ninesf}%
  \textfont\mibfam=\ninemib%
  \textfont\sybfam=\ninesyb%
  \bls{12pt}%
}

\def\tenpoint{
  \def\rm{\fam0\tenrm}%
  \textfont0=\tenrm \scriptfont0=\sevenrm \scriptscriptfont0=\fiverm%
  \textfont1=\teni  \scriptfont1=\seveni  \scriptscriptfont1=\fivei%
  \textfont2=\tensy \scriptfont2=\sevensy \scriptscriptfont2=\fivesy%
  \textfont\itfam=\tenit\def\it{\fam\itfam\tenit}%
  \textfont\bffam=\tenbf%
    \scriptfont\bffam=\sevenbf%
      \scriptscriptfont\bffam=\fivebf%
  \def\bf{\fam\bffam\tenbf}%
  \textfont\slfam=\tensl\def\sl{\fam\slfam\tensl}%
  \textfont\ttfam=\tentt\def\tt{\fam\ttfam\tentt}%
  \textfont\scfam=\tencsc\def\sc{\fam\scfam\tencsc}%
  \textfont\sffam=\tensf\def\sf{\fam\sffam\tensf}%
  \textfont\mibfam=\tenmib%
  \textfont\sybfam=\tensyb%
  \bls{12pt}%
}

\def\elevenpoint{
  \def\rm{\fam0\elevenrm}%
  \textfont0=\elevenrm \scriptfont0=\eightrm \scriptscriptfont0=\fiverm%
  \textfont1=\eleveni  \scriptfont1=\eighti  \scriptscriptfont1=\fivei%
  \textfont2=\elevensy \scriptfont2=\eightsy \scriptscriptfont2=\fivesy%
  \textfont\itfam=\elevenit\def\it{\fam\itfam\elevenit}%
  \textfont\bffam=\elevenbf%
    \scriptfont\bffam=\eightbf%
      \scriptscriptfont\bffam=\fivebf%
  \def\bf{\fam\bffam\elevenbf}%
  \textfont\slfam=\elevensl\def\sl{\fam\slfam\elevensl}%
  \textfont\ttfam=\eleventt\def\tt{\fam\ttfam\eleventt}%
  \textfont\scfam=\elevencsc\def\sc{\fam\scfam\elevencsc}%
  \textfont\sffam=\elevensf\def\sf{\fam\sffam\elevensf}%
  \textfont\mibfam=\elevenmib%
  \textfont\sybfam=\elevensyb%
  \bls{13pt}%
}

\def\fourteenpoint{
  \def\rm{\fam0\fourteenrm}%
  \textfont0\fourteenrm  \scriptfont0\tenrm  \scriptscriptfont0\sevenrm%
  \textfont1\fourteeni   \scriptfont1\teni   \scriptscriptfont1\seveni%
  \textfont2\fourteensy  \scriptfont2\tensy  \scriptscriptfont2\sevensy%
  \textfont\itfam=\fourteenit\def\it{\fam\itfam\fourteenit}%
  \textfont\bffam=\fourteenbf%
    \scriptfont\bffam=\tenbf%
      \scriptscriptfont\bffam=\sevenbf%
  \def\bf{\fam\bffam\fourteenbf}%
  \textfont\slfam=\fourteensl\def\sl{\fam\slfam\fourteensl}%
  \textfont\ttfam=\fourteentt\def\tt{\fam\ttfam\fourteentt}%
  \textfont\scfam=\fourteencsc\def\sc{\fam\scfam\fourteencsc}%
  \textfont\sffam=\fourteensf\def\sf{\fam\sffam\fourteensf}%
  \textfont\mibfam=\fourteenmib%
  \textfont\sybfam=\fourteensyb%
  \bls{17pt}%
}

\def\seventeenpoint{
  \def\rm{\fam0\seventeenrm}%
  \textfont0\seventeenrm  \scriptfont0\elevenrm  \scriptscriptfont0\ninerm%
  \textfont1\seventeeni   \scriptfont1\eleveni   \scriptscriptfont1\ninei%
  \textfont2\seventeensy  \scriptfont2\elevensy  \scriptscriptfont2\ninesy%
  \textfont\itfam=\seventeenit\def\it{\fam\itfam\seventeenit}%
  \textfont\bffam=\seventeenbf%
    \scriptfont\bffam=\elevenbf%
      \scriptscriptfont\bffam=\ninebf%
  \def\bf{\fam\bffam\seventeenbf}%
  \textfont\slfam=\seventeensl\def\sl{\fam\slfam\seventeensl}%
  \textfont\ttfam=\seventeentt\def\tt{\fam\ttfam\seventeentt}%
  \textfont\scfam=\seventeencsc\def\sc{\fam\scfam\seventeencsc}%
  \textfont\sffam=\seventeensf\def\sf{\fam\sffam\seventeensf}%
  \textfont\mibfam=\seventeenmib%
  \textfont\sybfam=\seventeensyb%
  \bls{20pt}%
}

\lineskip=1pt      \normallineskip=\lineskip
\lineskiplimit=0pt \normallineskiplimit=\lineskiplimit




\def\Nulle{0}  
\def\Aue{1}    
\def\Afe{2}    
\def\Sue{4}    
\def\Hae{5}    
\def\Hbe{6}    
\def\Hce{7}    
\def\Hde{8}    
\def\Kwe{9}    
\def\Txe{10}   
\def\Lie{11}   
\def\Bbe{12}   


\newdimen\DimenA
\newbox\BoxA

\newcount\LastMac \LastMac=\Nulle
\newcount\HeaderNumber \HeaderNumber=0
\newcount\DefaultHeader \DefaultHeader=\HeaderNumber
\newskip\Indent

\newskip\half      \half=5.5pt plus 1.5pt minus 2.25pt
\newskip\one       \one=11pt plus 3pt minus 5.5pt
\newskip\onehalf   \onehalf=16.5pt plus 5.5pt minus 8.25pt
\newskip\two       \two=22pt plus 5.5pt minus 11pt

\def\Half{\vskip-\lastskip\vskip\half}
\def\One{\vskip-\lastskip\vskip\one}
\def\OneHalf{\vskip-\lastskip\vskip\onehalf}
\def\Two{\vskip-\lastskip\vskip\two}


\def\rTenPT{10pt plus \Feathering}

\def\TenPT{10pt plus \Feathering} 
\def\ElevenPT{11pt plus \Feathering}

\def\Raggedright{
 \rightskip=0pt plus \hsize
}

\def\Fullout{
\rightskip=0pt
}

\def\Hang#1#2{
 \hangindent=#1
 \hangafter=#2
}

\def\EveryMac{
 \Fullout
 \everypar{}
}



\def\title#1{
 \EveryMac
 \LastMac=\Nulle
 \global\HeaderNumber=0
 \global\DefaultHeader=1
 \vbox to 1pc{\vss}
 \seventeenpoint
 \Raggedright
 \noindent \bf #1
}

\def\author#1{
 \EveryMac
 \ifnum\LastMac=\Afe \OneHalf
  \else \Two
 \fi
 \LastMac=\Aue
 \fourteenpoint
 \Raggedright
 \noindent \rm #1\par
 \vskip 3pt\relax
}

\def\affiliation#1{
 \EveryMac
 \LastMac=\Afe
 \eightpoint\bls{\TenPT}
 \Raggedright
 \noindent \it #1\par
}

\def\abstract{%
 \EveryMac
 \Two
 \LastMac=\Sue
 \everypar{\Hang{11pc}{0}}
 \noindent\ninebf ABSTRACT\par
 \tenpoint\bls{\ElevenPT}
 \Fullout
 \noindent\rm
}

\def\keywords{
 \EveryMac
 \Half
 \LastMac=\Kwe
 \everypar{\Hang{11pc}{0}}
 \tenpoint\bls{\ElevenPT}
 \Fullout
 \noindent\hbox{\bf Key words:\ }
 \rm
}


\def\maketitle{%
  \Two%
  \EndOpening%
  \MakePage%
}


\def\pageoffset#1#2{\hoffset=#1\relax\voffset=#2\relax}


\def\Autonumber{
 \global\AutoNumbertrue  
}

\newif\ifAutoNumber \AutoNumberfalse
\newcount\Sec        
\newcount\SecSec
\newcount\SecSecSec

\Sec=0

\def\:{\let\@sptoken= } \:  
\def\:{\@xifnch} \expandafter\def\: {\futurelet\@tempc\@ifnch}

\def\@ifnextchar#1#2#3{%
  \let\@tempMACe #1%
  \def\@tempMACa{#2}%
  \def\@tempMACb{#3}%
  \futurelet \@tempMACc\@ifnch%
}

\def\@ifnch{%
\ifx \@tempMACc \@sptoken%
  \let\@tempMACd\@xifnch%
\else%
  \ifx \@tempMACc \@tempMACe%
    \let\@tempMACd\@tempMACa%
  \else%
    \let\@tempMACd\@tempMACb%
  \fi%
\fi%
\@tempMACd%
}

\def\@ifstar#1#2{\@ifnextchar *{\def\@tempMACa*{#1}\@tempMACa}{#2}}

\def\section{\@ifstar{\@ssection}{\@section}}

\def\@section#1{
 \EveryMac
 \Two
 \LastMac=\Hae
 \ninepoint\bls{\ElevenPT}
 \bf
 \Raggedright
 \ifAutoNumber
  \advance\Sec by 1
  \noindent\number\Sec\hskip 1pc \uppercase{#1}
  \SecSec=0
 \else
  \noindent \uppercase{#1}
 \fi
 \nobreak
}

\def\@ssection#1{
 \EveryMac
 \ifnum\LastMac=\Hae \Half
  \else \OneHalf
 \fi
 \LastMac=\Hae
 \tenpoint\bls{\ElevenPT}
 \bf
 \Raggedright
 \noindent\uppercase{#1}
}

\def\subsection#1{
 \EveryMac
 \ifnum\LastMac=\Hae \Half
  \else \OneHalf
 \fi
 \LastMac=\Hbe
 \tenpoint\bls{\ElevenPT}
 \bf
 \Raggedright
 \ifAutoNumber
  \advance\SecSec by 1
  \noindent\number\Sec.\number\SecSec
  \hskip 1pc #1
  \SecSecSec=0
 \else
  \noindent #1
 \fi
 \nobreak
}

\def\subsubsection#1{
 \EveryMac
 \ifnum\LastMac=\Hbe \Half
  \else \OneHalf
 \fi
 \LastMac=\Hce
 \ninepoint\bls{\ElevenPT}
 \it
 \Raggedright
 \ifAutoNumber
  \advance\SecSecSec by 1
  \noindent\number\Sec.\number\SecSec.\number\SecSecSec
  \hskip 1pc #1
 \else
  \noindent #1
 \fi
 \nobreak
}

\def\paragraph#1{
 \EveryMac
 \One
 \LastMac=\Hde
 \ninepoint\bls{\ElevenPT}
 \noindent \it #1
 \rm
}


\def\tx{
 \EveryMac
 \ifnum\LastMac=\Lie \Half\fi
 \ifnum\LastMac=\Hae \nobreak\Half\fi
 \ifnum\LastMac=\Hbe \nobreak\Half\fi
 \ifnum\LastMac=\Hce \nobreak\Half\fi
 \ifnum\LastMac=\Lie \else \noindent\fi
 \LastMac=\Txe
 \ninepoint\bls{\ElevenPT}
 \rm
}


\def\item{
 \par
 \EveryMac
 \ifnum\LastMac=\Lie
  \else \Half
 \fi
 \LastMac=\Lie
 \ninepoint\bls{\ElevenPT}
 \rm
}


\def\bibitem{
 \par
 \EveryMac
 \ifnum\LastMac=\Bbe
  \else \Half
 \fi
 \LastMac=\Bbe
 \Hang{1.5em}{1}
 \eightpoint\bls{\TenPT}
 \Raggedright
 \noindent \rm
}


\newtoks\CatchLine

\def\@journal{Mon.\ Not.\ R.\ Astron.\ Soc.\ }  
\def\@pubyear{1995}        
\def\@pagerange{000--000}  
\def\@volume{000}          
\def\@microfiche{}         %

\def\pubyear#1{\gdef\@pubyear{#1}\@makecatchline}
\def\pagerange#1{\gdef\@pagerange{#1}\@makecatchline}
\def\volume#1{\gdef\@volume{#1}\@makecatchline}
\def\microfiche#1{\gdef\@microfiche{and Microfiche\ #1}\@makecatchline}

\def\@makecatchline{%
  \global\CatchLine{%
    {\rm \@journal {\bf \@volume},\ \@pagerange\ (\@pubyear)\ \@microfiche}}%
}

\@makecatchline 

\newtoks\LeftHeader
\def\shortauthor#1{
 \global\LeftHeader{#1}
}

\newtoks\RightHeader
\def\shorttitle#1{
 \global\RightHeader{#1}
}

\def\PageHead{
 \EveryMac
 \ifnum\HeaderNumber=1 \Pagehead
  \else \Catchline
 \fi
}

\def\Catchline{%
 \vbox to 0pt{\vskip-22.5pt
  \hbox to \PageWidth{\vbox to8.5pt{}\noindent
  \eightpoint\the\CatchLine\hfill}\vss}
 \nointerlineskip
}

\def\Pagehead{%
 \ifodd\pageno
   \vbox to 0pt{\vskip-22.5pt
   \hbox to \PageWidth{\vbox to8.5pt{}\elevenpoint\it\noindent
    \hfill\the\RightHeader\hskip1.5em\rm\folio}\vss}
 \else
   \vbox to 0pt{\vskip-22.5pt
   \hbox to \PageWidth{\vbox to8.5pt{}\elevenpoint\rm\noindent
   \folio\hskip1.5em\it\the\LeftHeader\hfill}\vss}
 \fi
 \nointerlineskip
}

\def\PageFoot{} 

\def\authorcomment#1{%
  \gdef\PageFoot{%
    \nointerlineskip%
    \vbox to 22pt{\vfil%
      \hbox to \PageWidth{\elevenpoint\rm\noindent \hfil #1 \hfil}}%
  }%
}

\everydisplay{\displaysetup}

\newif\ifeqno
\newif\ifleqno

\def\displaysetup#1$${%
 \displaytest#1\eqno\eqno\displaytest
}

\def\displaytest#1\eqno#2\eqno#3\displaytest{%
 \if!#3!\ldisplaytest#1\leqno\leqno\ldisplaytest
 \else\eqnotrue\leqnofalse\def\eqn{#2}\def\eq{#1}\fi
 \generaldisplay$$}

\def\ldisplaytest#1\leqno#2\leqno#3\ldisplaytest{%
 \def\eq{#1}%
 \if!#3!\eqnofalse\else\eqnotrue\leqnotrue
  \def\eqn{#2}\fi}

\def\generaldisplay{%
\ifeqno \ifleqno 
   \hbox to \hsize{\noindent
     $\displaystyle\eq$\hfil$\displaystyle\eqn$}
  \else
    \hbox to \hsize{\noindent
     $\displaystyle\eq$\hfil$\displaystyle\eqn$}
  \fi
 \else
 \hbox to \hsize{\vbox{\noindent
  $\displaystyle\eq$\hfil}}
 \fi
}

\def\@notice{%
  \par\Two%
  \bls{12pt}%
  \noindent\tenrm This paper has been produced using the Blackwell
                  Scientific Publications \TeX\ macros.%
}

\outer\def\bye{\@notice\par\vfill\supereject\end}

\everyjob{%
  \Warn{Monthly notices of the RAS journal style (\@typeface)\space
        v\@version,\space \@verdate.}\Warn{}%
}




\newif\if@debug \@debugfalse  

\def\Print#1{\if@debug\immediate\write16{#1}\else \fi}
\def\Warn#1{\immediate\write16{#1}}
\def\wlog#1{}

\newcount\Iteration 

\newif\ifFigureBoxes  
\FigureBoxestrue

\def\Single{0} \def\Double{1}                 
\def\Figure{0} \def\Table{1}                  

\def\InStack{0}  
\def\InZoneA{1}
\def\InZoneB{2}
\def\InZoneC{3}

\newcount\TEMPCOUNT 
\newdimen\TEMPDIMEN 
\newbox\TEMPBOX     
\newbox\VOIDBOX     

\newcount\LengthOfStack 
\newcount\MaxItems      
\newcount\StackPointer
\newcount\Point         
\newcount\NextFigure    
\newcount\NextTable     
\newcount\NextItem      

\newcount\StatusStack   
\newcount\NumStack      
\newcount\TypeStack     
\newcount\SpanStack     
\newcount\BoxStack      

\newcount\ItemSTATUS    
\newcount\ItemNUMBER    
\newcount\ItemTYPE      
\newcount\ItemSPAN      
\newbox\ItemBOX         
\newdimen\ItemSIZE      

\newdimen\PageHeight    
\newdimen\TextLeading   
\newdimen\Feathering    
\newcount\LinesPerPage  
\newdimen\ColumnWidth   
\newdimen\ColumnGap     
\newdimen\PageWidth     
\newdimen\BodgeHeight   
\newcount\Leading       

\newdimen\ZoneBSize  
\newdimen\TextSize   
\newbox\ZoneABOX     
\newbox\ZoneBBOX     
\newbox\ZoneCBOX     

\newif\ifFirstSingleItem
\newif\ifFirstZoneA
\newif\ifMakePageInComplete
\newif\ifMoreFigures \MoreFiguresfalse 
\newif\ifMoreTables  \MoreTablesfalse  

\newif\ifFigInZoneB 
\newif\ifFigInZoneC 
\newif\ifTabInZoneB 
\newif\ifTabInZoneC

\newif\ifZoneAFullPage

\newbox\MidBOX    
\newbox\LeftBOX
\newbox\RightBOX
\newbox\PageBOX   

\newif\ifLeftCOL  
\LeftCOLtrue

\newdimen\ZoneBAdjust

\newcount\ItemFits
\def\Yes{1}
\def\No{2}




\MaxItems=15
\NextFigure=0        
\NextTable=1

\BodgeHeight=6pt
\TextLeading=11pt    
\Leading=11
\Feathering=0pt      
\LinesPerPage=61     
\topskip=\TextLeading
\ColumnWidth=20pc    
\ColumnGap=2pc       

\def\ItemSep{\vskip \TextLeading plus \TextLeading minus 4pt}

\FigureBoxesfalse 

\parskip=0pt
\parindent=18pt
\widowpenalty=0
\clubpenalty=10000
\tolerance=1500
\hbadness=1500
\abovedisplayskip=6pt plus 2pt minus 2pt
\belowdisplayskip=6pt plus 2pt minus 2pt
\abovedisplayshortskip=6pt plus 2pt minus 2pt
\belowdisplayshortskip=6pt plus 2pt minus 2pt

\PageHeight=\TextLeading 
\multiply\PageHeight by \LinesPerPage
\advance\PageHeight by \topskip

\PageWidth=2\ColumnWidth
\advance\PageWidth by \ColumnGap




\newcount\DUMMY \StatusStack=\allocationnumber
\newcount\DUMMY \newcount\DUMMY \newcount\DUMMY 
\newcount\DUMMY \newcount\DUMMY \newcount\DUMMY 
\newcount\DUMMY \newcount\DUMMY \newcount\DUMMY
\newcount\DUMMY \newcount\DUMMY \newcount\DUMMY 
\newcount\DUMMY \newcount\DUMMY \newcount\DUMMY

\newcount\DUMMY \NumStack=\allocationnumber
\newcount\DUMMY \newcount\DUMMY \newcount\DUMMY 
\newcount\DUMMY \newcount\DUMMY \newcount\DUMMY 
\newcount\DUMMY \newcount\DUMMY \newcount\DUMMY 
\newcount\DUMMY \newcount\DUMMY \newcount\DUMMY 
\newcount\DUMMY \newcount\DUMMY \newcount\DUMMY

\newcount\DUMMY \TypeStack=\allocationnumber
\newcount\DUMMY \newcount\DUMMY \newcount\DUMMY 
\newcount\DUMMY \newcount\DUMMY \newcount\DUMMY 
\newcount\DUMMY \newcount\DUMMY \newcount\DUMMY 
\newcount\DUMMY \newcount\DUMMY \newcount\DUMMY 
\newcount\DUMMY \newcount\DUMMY \newcount\DUMMY

\newcount\DUMMY \SpanStack=\allocationnumber
\newcount\DUMMY \newcount\DUMMY \newcount\DUMMY 
\newcount\DUMMY \newcount\DUMMY \newcount\DUMMY 
\newcount\DUMMY \newcount\DUMMY \newcount\DUMMY 
\newcount\DUMMY \newcount\DUMMY \newcount\DUMMY 
\newcount\DUMMY \newcount\DUMMY \newcount\DUMMY

\newbox\DUMMY   \BoxStack=\allocationnumber
\newbox\DUMMY   \newbox\DUMMY \newbox\DUMMY 
\newbox\DUMMY   \newbox\DUMMY \newbox\DUMMY 
\newbox\DUMMY   \newbox\DUMMY \newbox\DUMMY 
\newbox\DUMMY   \newbox\DUMMY \newbox\DUMMY 
\newbox\DUMMY   \newbox\DUMMY \newbox\DUMMY

\def\wlog{\immediate\write-1}


\def\GetItemAll#1{%
 \GetItemSTATUS{#1}
 \GetItemNUMBER{#1}
 \GetItemTYPE{#1}
 \GetItemSPAN{#1}
 \GetItemBOX{#1}
}

\def\GetItemSTATUS#1{%
 \Point=\StatusStack
 \advance\Point by #1
 \global\ItemSTATUS=\count\Point
}

\def\GetItemNUMBER#1{%
 \Point=\NumStack
 \advance\Point by #1
 \global\ItemNUMBER=\count\Point
}

\def\GetItemTYPE#1{%
 \Point=\TypeStack
 \advance\Point by #1
 \global\ItemTYPE=\count\Point
}

\def\GetItemSPAN#1{%
 \Point\SpanStack
 \advance\Point by #1
 \global\ItemSPAN=\count\Point
}

\def\GetItemBOX#1{%
 \Point=\BoxStack
 \advance\Point by #1
 \global\setbox\ItemBOX=\vbox{\copy\Point}
 \global\ItemSIZE=\ht\ItemBOX
 \global\advance\ItemSIZE by \dp\ItemBOX
 \TEMPCOUNT=\ItemSIZE
 \divide\TEMPCOUNT by \Leading
 \divide\TEMPCOUNT by 65536
 \advance\TEMPCOUNT by 1
 \ItemSIZE=\TEMPCOUNT pt
 \global\multiply\ItemSIZE by \Leading
}


\def\JoinStack{%
 \ifnum\LengthOfStack=\MaxItems 
  \Warn{WARNING: Stack is full...some items will be lost!}
 \else
  \Point=\StatusStack
  \advance\Point by \LengthOfStack
  \global\count\Point=\ItemSTATUS
  \Point=\NumStack
  \advance\Point by \LengthOfStack
  \global\count\Point=\ItemNUMBER
  \Point=\TypeStack
  \advance\Point by \LengthOfStack
  \global\count\Point=\ItemTYPE
  \Point\SpanStack
  \advance\Point by \LengthOfStack
  \global\count\Point=\ItemSPAN
  \Point=\BoxStack
  \advance\Point by \LengthOfStack
  \global\setbox\Point=\vbox{\copy\ItemBOX}
  \global\advance\LengthOfStack by 1
  \ifnum\ItemTYPE=\Figure 
   \global\MoreFigurestrue
  \else
   \global\MoreTablestrue
  \fi
 \fi
}


\def\LeaveStack#1{%
 {\Iteration=#1
 \loop
 \ifnum\Iteration<\LengthOfStack
  \advance\Iteration by 1
  \GetItemSTATUS{\Iteration}
   \advance\Point by -1
   \global\count\Point=\ItemSTATUS
  \GetItemNUMBER{\Iteration}
   \advance\Point by -1
   \global\count\Point=\ItemNUMBER
  \GetItemTYPE{\Iteration}
   \advance\Point by -1
   \global\count\Point=\ItemTYPE
  \GetItemSPAN{\Iteration}
   \advance\Point by -1
   \global\count\Point=\ItemSPAN
  \GetItemBOX{\Iteration}
   \advance\Point by -1
   \global\setbox\Point=\vbox{\copy\ItemBOX}
 \repeat}
 \global\advance\LengthOfStack by -1
}


\newif\ifStackNotClean

\def\CleanStack{%
 \StackNotCleantrue
 {\Iteration=0
  \loop
   \ifStackNotClean
    \GetItemSTATUS{\Iteration}
    \ifnum\ItemSTATUS=\InStack
     \advance\Iteration by 1
     \else
      \LeaveStack{\Iteration}
    \fi
   \ifnum\LengthOfStack<\Iteration
    \StackNotCleanfalse
   \fi
 \repeat}
}


\def\FindItem#1#2{%
 \global\StackPointer=-1 
 {\Iteration=0
  \loop
  \ifnum\Iteration<\LengthOfStack
   \GetItemSTATUS{\Iteration}
   \ifnum\ItemSTATUS=\InStack
    \GetItemTYPE{\Iteration}
    \ifnum\ItemTYPE=#1
     \GetItemNUMBER{\Iteration}
     \ifnum\ItemNUMBER=#2
      \global\StackPointer=\Iteration
      \Iteration=\LengthOfStack 
     \fi
    \fi
   \fi
  \advance\Iteration by 1
 \repeat}
}


\def\FindNext{%
 \global\StackPointer=-1 
 {\Iteration=0
  \loop
  \ifnum\Iteration<\LengthOfStack
   \GetItemSTATUS{\Iteration}
   \ifnum\ItemSTATUS=\InStack
    \GetItemTYPE{\Iteration}
   \ifnum\ItemTYPE=\Figure
    \ifMoreFigures
      \global\NextItem=\Figure
      \global\StackPointer=\Iteration
      \Iteration=\LengthOfStack 
    \fi
   \fi
   \ifnum\ItemTYPE=\Table
    \ifMoreTables
      \global\NextItem=\Table
      \global\StackPointer=\Iteration
      \Iteration=\LengthOfStack 
    \fi
   \fi
  \fi
  \advance\Iteration by 1
 \repeat}
}


\def\ChangeStatus#1#2{%
 \Point=\StatusStack
 \advance\Point by #1
 \global\count\Point=#2
}



\def\Zone{\InZoneA}

\ZoneBAdjust=0pt

\def\MakePage{
 \global\ZoneBSize=\PageHeight
 \global\TextSize=\ZoneBSize
 \global\ZoneAFullPagefalse
 \global\topskip=\TextLeading
 \MakePageInCompletetrue
 \MoreFigurestrue
 \MoreTablestrue
 \FigInZoneBfalse
 \FigInZoneCfalse
 \TabInZoneBfalse
 \TabInZoneCfalse
 \global\FirstSingleItemtrue
 \global\FirstZoneAtrue
 \global\setbox\ZoneABOX=\box\VOIDBOX
 \global\setbox\ZoneBBOX=\box\VOIDBOX
 \global\setbox\ZoneCBOX=\box\VOIDBOX
 \loop
  \ifMakePageInComplete
 \FindNext
 \ifnum\StackPointer=-1
  \NextItem=-1
  \MoreFiguresfalse
  \MoreTablesfalse
 \fi
 \ifnum\NextItem=\Figure
   \FindItem{\Figure}{\NextFigure}
   \ifnum\StackPointer=-1 \global\MoreFiguresfalse
   \else
    \GetItemSPAN{\StackPointer}
    \ifnum\ItemSPAN=\Single \def\Zone{\InZoneB}\relax
     \ifFigInZoneC \global\MoreFiguresfalse\fi
    \else
     \def\Zone{\InZoneA}
     \ifFigInZoneB \def\Zone{\InZoneC}\fi
    \fi
   \fi
   \ifMoreFigures\Print{}\FigureItems\fi
 \fi
\ifnum\NextItem=\Table
   \FindItem{\Table}{\NextTable}
   \ifnum\StackPointer=-1 \global\MoreTablesfalse
   \else
    \GetItemSPAN{\StackPointer}
    \ifnum\ItemSPAN=\Single\relax
     \ifTabInZoneC \global\MoreTablesfalse\fi
    \else
     \def\Zone{\InZoneA}
     \ifTabInZoneB \def\Zone{\InZoneC}\fi
    \fi
   \fi
   \ifMoreTables\Print{}\TableItems\fi
 \fi
   \MakePageInCompletefalse 
   \ifMoreFigures\MakePageInCompletetrue\fi
   \ifMoreTables\MakePageInCompletetrue\fi
 \repeat
 \ifZoneAFullPage
  \global\TextSize=0pt
  \global\ZoneBSize=0pt
  \global\vsize=0pt\relax
  \global\topskip=0pt\relax
  \vbox to 0pt{\vss}
  \eject
 \else
 \global\advance\ZoneBSize by -\ZoneBAdjust
 \global\vsize=\ZoneBSize
 \global\hsize=\ColumnWidth
 \global\ZoneBAdjust=0pt
 \ifdim\TextSize<23pt
 \Warn{}
 \Warn{* Making column fall short: TextSize=\the\TextSize *}
 \vskip-\lastskip\eject\fi
 \fi
}

\def\MakeRightCol{
 \global\TextSize=\ZoneBSize
 \MakePageInCompletetrue
 \MoreFigurestrue
 \MoreTablestrue
 \global\FirstSingleItemtrue
 \global\setbox\ZoneBBOX=\box\VOIDBOX
 \def\Zone{\InZoneB}
 \loop
  \ifMakePageInComplete
 \FindNext
 \ifnum\StackPointer=-1
  \NextItem=-1
  \MoreFiguresfalse
  \MoreTablesfalse
 \fi
 \ifnum\NextItem=\Figure
   \FindItem{\Figure}{\NextFigure}
   \ifnum\StackPointer=-1 \MoreFiguresfalse
   \else
    \GetItemSPAN{\StackPointer}
    \ifnum\ItemSPAN=\Double\relax
     \MoreFiguresfalse\fi
   \fi
   \ifMoreFigures\Print{}\FigureItems\fi
 \fi
 \ifnum\NextItem=\Table
   \FindItem{\Table}{\NextTable}
   \ifnum\StackPointer=-1 \MoreTablesfalse
   \else
    \GetItemSPAN{\StackPointer}
    \ifnum\ItemSPAN=\Double\relax
     \MoreTablesfalse\fi
   \fi
   \ifMoreTables\Print{}\TableItems\fi
 \fi
   \MakePageInCompletefalse 
   \ifMoreFigures\MakePageInCompletetrue\fi
   \ifMoreTables\MakePageInCompletetrue\fi
 \repeat
 \ifZoneAFullPage
  \global\TextSize=0pt
  \global\ZoneBSize=0pt
  \global\vsize=0pt\relax
  \global\topskip=0pt\relax
  \vbox to 0pt{\vss}
  \eject
 \else
 \global\vsize=\ZoneBSize
 \global\hsize=\ColumnWidth
 \ifdim\TextSize<23pt
 \Warn{}
 \Warn{* Making column fall short: TextSize=\the\TextSize *}
 \vskip-\lastskip\eject\fi
\fi
}

\def\FigureItems{
 \Print{Considering...}
 \ShowItem{\StackPointer}
 \GetItemBOX{\StackPointer} 
 \GetItemSPAN{\StackPointer}
  \CheckFitInZone 
  \ifnum\ItemFits=\Yes
   \ifnum\ItemSPAN=\Single
     \ChangeStatus{\StackPointer}{\InZoneB} 
     \global\FigInZoneBtrue
     \ifFirstSingleItem
      \hbox{}\vskip-\BodgeHeight
     \global\advance\ItemSIZE by \TextLeading
     \fi
     \unvbox\ItemBOX\ItemSep
     \global\FirstSingleItemfalse
     \global\advance\TextSize by -\ItemSIZE
     \global\advance\TextSize by -\TextLeading
   \else
    \ifFirstZoneA
     \global\advance\ItemSIZE by \TextLeading
     \global\FirstZoneAfalse\fi
    \global\advance\TextSize by -\ItemSIZE
    \global\advance\TextSize by -\TextLeading
    \global\advance\ZoneBSize by -\ItemSIZE
    \global\advance\ZoneBSize by -\TextLeading
    \ifFigInZoneB\relax
     \else
     \ifdim\TextSize<3\TextLeading
     \global\ZoneAFullPagetrue
     \fi
    \fi
    \ChangeStatus{\StackPointer}{\Zone}
    \ifnum\Zone=\InZoneC \global\FigInZoneCtrue\fi
  \fi
   \Print{TextSize=\the\TextSize}
   \Print{ZoneBSize=\the\ZoneBSize}
  \global\advance\NextFigure by 1
   \Print{This figure has been placed.}
  \else
   \Print{No space available for this figure...holding over.}
   \Print{}
   \global\MoreFiguresfalse
  \fi
}

\def\TableItems{
 \Print{Considering...}
 \ShowItem{\StackPointer}
 \GetItemBOX{\StackPointer} 
 \GetItemSPAN{\StackPointer}
  \CheckFitInZone 
  \ifnum\ItemFits=\Yes
   \ifnum\ItemSPAN=\Single
    \ChangeStatus{\StackPointer}{\InZoneB}
     \global\TabInZoneBtrue
     \ifFirstSingleItem
      \hbox{}\vskip-\BodgeHeight
     \global\advance\ItemSIZE by \TextLeading
     \fi
     \unvbox\ItemBOX\ItemSep
     \global\FirstSingleItemfalse
     \global\advance\TextSize by -\ItemSIZE
     \global\advance\TextSize by -\TextLeading
   \else
    \ifFirstZoneA
    \global\advance\ItemSIZE by \TextLeading
    \global\FirstZoneAfalse\fi
    \global\advance\TextSize by -\ItemSIZE
    \global\advance\TextSize by -\TextLeading
    \global\advance\ZoneBSize by -\ItemSIZE
    \global\advance\ZoneBSize by -\TextLeading
    \ifFigInZoneB\relax
     \else
     \ifdim\TextSize<3\TextLeading
     \global\ZoneAFullPagetrue
     \fi
    \fi
    \ChangeStatus{\StackPointer}{\Zone}
    \ifnum\Zone=\InZoneC \global\TabInZoneCtrue\fi
   \fi
  \global\advance\NextTable by 1
   \Print{This table has been placed.}
  \else
  \Print{No space available for this table...holding over.}
   \Print{}
   \global\MoreTablesfalse
  \fi
}


\def\CheckFitInZone{%
{\advance\TextSize by -\ItemSIZE
 \advance\TextSize by -\TextLeading
 \ifFirstSingleItem
  \advance\TextSize by \TextLeading
 \fi
 \ifnum\Zone=\InZoneA\relax
  \else \advance\TextSize by -\ZoneBAdjust
 \fi
 \ifdim\TextSize<3\TextLeading \global\ItemFits=\No
 \else \global\ItemFits=\Yes\fi}
}

\def\BF#1#2{
 \ItemSTATUS=\InStack
 \ItemNUMBER=#1
 \ItemTYPE=\Figure
 \if#2S \ItemSPAN=\Single
  \else \ItemSPAN=\Double
 \fi
 \setbox\ItemBOX=\vbox{}
}

\def\BT#1#2{
 \ItemSTATUS=\InStack
 \ItemNUMBER=#1
 \ItemTYPE=\Table
 \if#2S \ItemSPAN=\Single
  \else \ItemSPAN=\Double
 \fi
 \setbox\ItemBOX=\vbox{}
}

\def\BeginOpening{%
 \hsize=\PageWidth
 \global\setbox\ItemBOX=\vbox\bgroup
}

\let\begintopmatter=\BeginOpening  

\def\EndOpening{%
 \egroup
 \ItemNUMBER=0
 \ItemTYPE=\Figure
 \ItemSPAN=\Double
 \ItemSTATUS=\InStack
 \JoinStack
}


\newbox\tmpbox

\def\FC#1#2#3#4{%
  \ItemSTATUS=\InStack
  \ItemNUMBER=#1
  \ItemTYPE=\Figure
  \if#2S
    \ItemSPAN=\Single \TEMPDIMEN=\ColumnWidth
  \else
    \ItemSPAN=\Double \TEMPDIMEN=\PageWidth
  \fi
  {\hsize=\TEMPDIMEN
   \global\setbox\ItemBOX=\vbox{%
     \ifFigureBoxes
       \B{\TEMPDIMEN}{#3}
     \else
       \vbox to #3{\vfil}%
     \fi%
     \eightpoint\rm\bls{\rTenPT}%
     \vskip 5.5pt plus 6pt%
     \setbox\tmpbox=\vbox{#4\par}%
     \ifdim\ht\tmpbox>10pt 
       \noindent #4\par%
     \else
       \hbox to \hsize{\hfil #4\hfil}%
     \fi%
   }%
  }%
  \JoinStack%
  \Print{Processing source for figure {\the\ItemNUMBER}}%
}


\def\TH#1#2#3#4{%
 \ItemSTATUS=\InStack
 \ItemNUMBER=#1
 \ItemTYPE=\Table
 \if#2S \ItemSPAN=\Single \TEMPDIMEN=\ColumnWidth
  \else \ItemSPAN=\Double \TEMPDIMEN=\PageWidth
 \fi
{\hsize=\TEMPDIMEN
\eightpoint\bls{\rTenPT}\rm
\global\setbox\ItemBOX=\vbox{\noindent#3\vskip 5.5pt plus5.5pt\noindent#4}}
 \JoinStack
 \Print{Processing source for table {\the\ItemNUMBER}}
}

\let\table=\TH  

\def\UnloadZoneA{%
\FirstZoneAtrue
 \Iteration=0
  \loop
   \ifnum\Iteration<\LengthOfStack
    \GetItemSTATUS{\Iteration}
    \ifnum\ItemSTATUS=\InZoneA
     \GetItemBOX{\Iteration}
     \ifFirstZoneA \vbox to \BodgeHeight{\vfil}%
     \FirstZoneAfalse\fi
     \unvbox\ItemBOX\ItemSep
     \LeaveStack{\Iteration}
     \else
     \advance\Iteration by 1
   \fi
 \repeat
}

\def\UnloadZoneC{%
\Iteration=0
  \loop
   \ifnum\Iteration<\LengthOfStack
    \GetItemSTATUS{\Iteration}
    \ifnum\ItemSTATUS=\InZoneC
     \GetItemBOX{\Iteration}
     \ItemSep\unvbox\ItemBOX
     \LeaveStack{\Iteration}
     \else
     \advance\Iteration by 1
   \fi
 \repeat
}


\def\ShowItem#1{
  {\GetItemAll{#1}
  \Print{\the#1:
  {TYPE=\ifnum\ItemTYPE=\Figure Figure\else Table\fi}
  {NUMBER=\the\ItemNUMBER}
  {SPAN=\ifnum\ItemSPAN=\Single Single\else Double\fi}
  {SIZE=\the\ItemSIZE}}}
}

\def\ShowStack{%
 \Print{}
 \Print{LengthOfStack = \the\LengthOfStack}
 \ifnum\LengthOfStack=0 \Print{Stack is empty}\fi
 \Iteration=0
 \loop
 \ifnum\Iteration<\LengthOfStack
  \ShowItem{\Iteration}
  \advance\Iteration by 1
 \repeat
}

\def\B#1#2{%
\hbox{\vrule\kern-0.4pt\vbox to #2{%
\hrule width #1\vfill\hrule}\kern-0.4pt\vrule}
}

\def\Ref#1{\begingroup\global\setbox\TEMPBOX=\vbox{\hsize=2in\noindent#1}\endgroup
\ht1=0pt\dp1=0pt\wd1=0pt\vadjust{\vtop to 0pt{\advance
\hsize0.5pc\kern-10pt\moveright\hsize\box\TEMPBOX\vss}}}

\def\MarkRef#1{\leavevmode\thinspace\hbox{\vrule\vtop
{\vbox{\hrule\kern1pt\hbox{\vphantom{\rm/}\thinspace{\rm#1}%
\thinspace}}\kern1pt\hrule}\vrule}\thinspace}%


\output{%
 \ifLeftCOL
  \global\setbox\LeftBOX=\vbox to \ZoneBSize{\box255\unvbox\ZoneBBOX}
  \global\LeftCOLfalse
  \MakeRightCol
 \else
  \setbox\RightBOX=\vbox to \ZoneBSize{\box255\unvbox\ZoneBBOX}
  \setbox\MidBOX=\hbox{\box\LeftBOX\hskip\ColumnGap\box\RightBOX}
  \setbox\PageBOX=\vbox to \PageHeight{%
  \UnloadZoneA\box\MidBOX\UnloadZoneC}
  \shipout\vbox{\PageHead\box\PageBOX\PageFoot}
  \global\advance\pageno by 1
  \global\HeaderNumber=\DefaultHeader
  \global\LeftCOLtrue
  \CleanStack
  \MakePage
 \fi
}


\catcode `\@=12 


\def\ergcms{erg$\,$s$^{-1}$cm$^{-2}$}

\pageoffset{-2pc}{0pc}

\def\A{\AA}

\def\ergscm2{ergs cm$^{-2}$ s$^{-1}$}
\def\ergcms{ergs cm$^{-2}$ s$^{-1}$}
\def\ergcm2{ergs cm$^{-2}$ s$^{-1}$}

\def\laeq{\lower.5ex\hbox{\fiverm{\ $\buildrel < \over \sim$\ }}}
\def\gaeq{\lower.5ex\hbox{\fiverm{\ $\buildrel > \over \sim$\ }}}
\def\boxit#1{\vbox{\hrule\hbox{\vrule\kern3pt
     \vbox{\kern3pt#1\kern3pt}\kern3pt\vrule}\hrule}}

\def\BJBa4{
 \hsize=6.3in
 \vsize=9.6in
 \voffset=-0.3in
}

\def\ecs{ergs cm$^{-2}$ sec$^{-1}$}
\def\es{ergs sec$^{-1}$}

\def\ref{\noindent} 

\def\gaeq{$_ >\atop{^\sim}$}
\def\laeq{$_ <\atop{^\sim}$}
\def\ergcms{erg$\,$s$^{-1}$cm$^{-2}$}


\pageoffset{-2pc}{0pc}

\Autonumber
\begintopmatter

\title{A Deep ROSAT Survey X: X-ray Luminous Narrow Emission Line Galaxies}
\author{R.E.Griffiths,$^1$ 
R.Della Ceca,$^{1}$ I.Georgantopoulos,$^2$ B.J.Boyle,$^3$  G.C.Stewart$^2$, T.Shanks,$^4$ and A.Fruscione$^5$}
\affiliation{1. Department of Physics and Astronomy, Johns
Hopkins University,  Homewood Campus, Baltimore, MD 21218, USA\hfil\break
 2. Department of Physics and Astronomy, The University, University Road,
Leicester, LE1 7RH\hfil\break
3. Institute of Astronomy, University of Cambridge, Madingley
Road, Cambridge, CB3 0HA\hfil\break
 4. Department of Physics,
University of Durham, South Road,  Durham, DH1 3LE\hfill\break
 5. Center for EUV Astrophysics, 2150 Kittredge Street, University of 
California, Berkeley, CA 94720}

\shortauthor{R.E.Griffiths et al.}
\shorttitle{Narrow Emission Line Galaxies}

\abstract
X-ray luminous narrow emision-line galaxies (NELG) have been previously
identified and proposed as an important class of extragalactic X-ray
source, with a potentially significant contribution to the total
extragalactic X-ray flux at energies below $\sim$ 10 keV.  In
order to investigate and clarify this possibility, we have used a sample
of NELG found in 5 deep ROSAT fields and  similar samples belonging to the
Cambridge-Cambridge ROSAT Serendipity Survey  and to the 
{\it Einstein} Observatory Extended Medium Sensitivity Survey sample.
We have been able to study their X-ray properties, to derive their
number-flux relationship, to investigate their cosmological evolution
and to derive a preliminary X-ray luminosity function (XLF) for this 
class of objects.

We have compared the above mentioned properties to those exhibited by
soft X-ray selected Broad Line AGN (BLAGN) and/or normal galaxies.  
The principal results of
this investigation are as follows: a) for a given optical luminosity,
the typical X-ray luminosity of NELGs is about one or two
orders of magnitude higher than that of normal galaxies; b) the
ratio of the surface density of NELGs compared with BLAGN increases
from $\sim 0.04$ at $f_x$ \gaeq $6\times 10^{-13}$ \ecs to $\sim $ 0.1 at
$f_x$ \gaeq$10^{-14}$ \ecs, suggesting that the surface density of NL
galaxies might be very close to that of BLAGN at $f_x \sim 10^{-15}$ \ecs;
c) we find that these objects are described by a cosmological evolution
rate similar to that of soft X-ray selected BLAGN; d) the de-evolved 
($z=0$) XLF of NELGs in the luminosity range $10^{41} - 5\times 10^{43}$ \es
is steeper than the BLAGN ($z=0$) XLF in the same luminosity interval.  
Their
spatial density is significantly lower than the spatial density of X-ray
selected BLAGN at $L_x(z=0) \sim 5\times 10^{43}$ \es, but this difference
decreases at lower luminosities such that at $L_x(z=0) \le 10^{42}$ \es the
spatial density of NELGs is very close to that of BLAGN.  The
implications of these results for the contribution of this class of
objects to the cosmic X-ray background are discussed.

\keywords X-rays: general -- galaxies: active -- galaxies:Seyfert --
galaxies:starburst -- X-ray:galaxies -- cosmology

\maketitle

\vfil\eject

\section{Introduction}\tx

The well established classes of extragalactic X-ray source comprise (i)
groups and clusters of galaxies, (ii) active galactic nuclei (AGN) and
(iii) ``normal'' galaxies. For a review of these topics, see de Zotti et
al. (1995) and papers in Barcons and Fabian (1992).  One fork of the
high-luminosity end of the ``normal'' galaxy luminosity function
comprises the ``starburst'' galaxies. At least one example of an X-ray
emitting irregular starburst galaxy has been known for two decades,
viz. M82 (catalogued in the UHURU and ARIEL V sky surveys). The relative
importance of the class of objects typified by M82 has, however, been
unclear.  Although a handful of ``narrow-line '' galaxies were
identified from the early SAS-3 and HEAO-A3 data (e.g. Schnopper et al,
1978, Griffiths et al 1979), it later became clear that these were
predominantly Seyfert 2 galaxies.
The suggestion that there might be a class of X-ray luminous
``starburst'' galaxies was first made by Weedman (1986), based on
Einstein observations of a small number of optically selected peculiar
or irregular galaxies (Fabbiano, Feigelson and Zamorani 1982) for which
IRAS fluxes were later obtained. An observed correlation between X-ray
and infrared luminosities of an expanded sample of ``starburst''
galaxies led several authors (Griffiths and Padovani, 1990 (GP90);
Green, Anderson and Ward, 1992; David, Jones and Forman, 1992; Treyer et
al 1992) to point out that such galaxies might represent an important
class of extragalactic X-ray source, with a potentially significant
contribution to the total extragalactic X-ray ``background'' (XRB) at
the level of several tens of percent at 2 keV, and perhaps a similar
contribution up to at least 10 or even 20 keV.

\bigskip
There were two possibilities summarized by GP90; (i) X-ray emission from
luminous infrared galaxies presumably powered by central or distributed
starburst activity, and (ii) X-ray emission from blue, star-forming
galaxies of the kind contributing to the faint blue number counts. The
latter category includes dwarf galaxies with low metallicity (e.g.  NGC
5408 -- Stewart et al. (1982), and low-metallicity starburst
regions within spiral or irregular galaxies.  Although there is probably
a considerable overlap between the galaxies in these two broad
categories, it may be important to try to separate their optical or
infrared properties in terms of correlations with X-ray emission.
In order to confirm the predictions of the Einstein/IRAS galaxy
correlation, Fruscione and Griffiths (1991) and Fruscione, Griffiths and
Mackenty (1993) performed low resolution optical spectroscopy (in the
range $3500-7300\AA$) of galaxies selected from the {\it Einstein}
Observatory Extended Medium Sensitivity Survey (EMSS -- see Gioia et
al. 1990, Stocke et al. 1991) to establish the existence of X-ray
luminous starburst galaxies, and found 6 sources.

\bigskip
In order to investigate and clarify the nature of sources near the ROSAT deep
survey limit, we have performed independent $\sim 30-60$ ks ROSAT deep
X-ray surveys of 5 fields, which had previously been surveyed for
UV-excess quasars by Boyle et al. (1990).  About ten X-ray sources in
each field were immediately identifiable with UV-excess quasars, and the
classification and properties of the remaining sources have been pursued
via optical spectroscopy of the corresponding candidate counterparts
At the flux limit (0.5 - 2.0 keV) of
$\sim 4\times 10^{-15}$ \ecs our optical spectroscopy has shown that the
majority of these sources are BLAGN at a mean redshift of $z \sim$ 1.5
(Shanks et al., 1991; Boyle et al., 1993;1994), with an important
minority ($\sim 10\%$) of X-ray luminous galaxies, consisting of both
early-type ``passive" galaxies and late-type Emission Line Galaxies
(Griffiths et al 1995).
The BLAGN content of these fields has been described by Boyle et al (1993,
1994), who derived the BLAGN luminosity function and the constraints on its
evolution.  Based on the observed range in the parameter values for the
best-fit evolutionary models, BLAGN can account for up to $55\%$ of the
X-ray background between 1 and 2 keV, i.e. they certainly do {\it not}
account for all the observed soft X-ray background.  Similar conclusions
were reached by Georgantopoulus et al., (1993, 1995), 
 using arguments based on the BLAGN log(N)-log(S), on the
comparison between the BLAGN spectral properties and the XRB spectral
properties, and on the auto-correlation analysis of the XRB.

\bigskip
The above arguments suggest the existence of at least one new population
contributing substantially to the soft XRB (Georgantopoulos et al 1995),
and from our deep survey observations we explore the possibility that this
new population could be comprised of X-ray luminous galaxies.  This
suggestion has been supported by the detection of a strong ($> 5\sigma$)
signal below 10$^{\prime\prime}$ in the cross-correlation function of
the $< 4\sigma$ X-ray background fluctuations with B $<$ 23 galaxies
(Roche, et al., 1995).  These considerations have thus led us to
investigate the non-BLAGN counterparts to the ROSAT deep survey sources,
and, in particular, the NELG content. An investigation of the properties
of NELGs may provide a useful constraint on their cosmological
evolution.  Unfortunately, with the spectroscopic data at our disposal
we can not classify our NELGs as starburst-like or Seyfert 2 like
objects using the standard line ratio tests of Veilleux and Osterbrock
(1987).  Based on high resolution, high signal-to-noise optical
spectroscopy of similar objects found in the Cambridge-Cambridge ROSAT
Serendipity Survey (CRSS) sample (Boyle et al 1995a,b) and in the EMSS
sample (see below) we can conjecture that our NELGs are a mix of
starburst, Seyfert II and composite galaxies.

The possible connections between AGN and starburst nuclei have been
discussed by Norman and Scoville (1988), Heckman (1991), Williams and
Perry (1994), Filippenko (1992) and others. It has been demonstrated
that in some nearby objects such as NGC1068 (Wilson et al. 1992), NGC
2992 (Elvis et al. 1992), NGC1672, and NGC1566, composite nuclear
regions have been found where there is evidence of both starburst and
Seyfert 2 activity.  Using infrared observations, Maiolino et al. (1995)
have pointed out that Seyfert 2 nuclei tend to lie in galaxies
experiencing enhanced star-forming activity (unlike Seyfert 1 nuclei,
which lie in galaxies with normal or slightly enhanced nuclear star
formation). We note that a population of narrow-line Seyfert 1 galaxies
with soft X-ray spectra has been identified using the ROSAT all-sky
survey data (Boller, Brandt and Fink 1995), but but these have extremely
large FeII/H$\beta$ ratios and are spectroscopically different from the objects
discussed here.

With the close connection between nuclear starbursts and Seyfert 2
nuclei, it may be appropriate to consider them for the present purposees
as a joint sample, in the absence of clear observational evidence for
separation of these populations. In this vein, Boyle et al., (1995a)
used a combined sample of 43 objects composed of (a) 12 Narrow Emission
Line Galaxies found at $f_x$ (0.5 - 2.0 keV) $> \ 2\times 10^{-14}$ \ecs
in the CRSS and (b) 31 ``ambiguous'' EMSS sources. The combined sample
was used to constraint the cosmological evolution of starbursts/Seyfert
II galaxies.  In this paper we have been able to make two major
improvements over the work of Boyle et al.  First, we increase the
number of NELGs found in the ROSAT deep surveys at faint fluxes ($f_x$
(0.5 - 2.0 keV) \gaeq $4\times 10^{-15}$ \ecs ) by about $60\%$. Our
flux limit is almost a factor 5 fainter than the CRSS flux limit.
Second, using spectroscopic observations of 15 EMSS ``ambiguous''
sources, along with other data from the literature, we have been able to
substantially improve our knowlege of the NELGs in the EMSS.

\bigskip
This paper is organized as follows: in section II we present and discuss
our working data set.  Having spectroscopically identified $\sim 75\%$
of the ROSAT sources in our fields, we also discuss the distribution of
these sources in the ($f_x/f_B$) $-$ $f_B$ plane. As previously noted by
Gioia et al. (1984) and Stocke et al. (1991), this information is
extremely useful in programmes of spectroscopic identification of faint
X-ray sources.  We also report new spectroscopic results
on 15 ``ambiguous'' sources belonging to the EMSS sample.  In section
III we derive the X-ray number-flux relationship of NELG
and compare it with the BLAGN Log N($>S$) - Log S obtained from the
combined EMSS-ROSAT sample.  We also compare the derived Log N($>S$) -
Log S with the predictions of GP90 which were
based on an infrared sample of starburst galaxies and on the measured
relation between infrared and X-ray luminosity.  In section IV we
investigate the cosmological evolution of the NELG,
derive an X-ray luminosity function for this class of objects
and discuss their contribution to the soft X-ray background.  Finally in
section V we present our summary and conclusions.
\bigskip
 
A Hubble constant of 50 $km\ s^{-1}$ and a Friedmann universe with 
a deceleration parameter $q_o = 0$ are assumed throughout.

\vfill\eject

\section {The data}\tx
\noindent

\subsection{Our ROSAT Sample of Narrow Line Galaxies }\tx
\noindent

Our X-ray sample is based on 5 deep exposures taken with the Position
Sensitive Proportional Counter (PSPC; Pfeffermann et al., 1986) onboard
the ROSAT satellite (Truemper 1983).  These fields were
previously surveyed spectroscopically for QSOs as part of the Durham UVX
QSO survey by Boyle et al. (1990).  The names, field centres and exposure
times for each PSPC field are reported in Table 1 of Boyle et al. (1994).
Full details of the X-ray and optical observations are presented
elsewhere (Boyle et al. (1994), Georgantopoulus et al. (1995) ; Shanks et
al., in preparation) and thus only brief details will be given here.
 
One hundred and ninety four sources were detected above the 
$5 \sigma$ detection threshold in the (0.5-2.0) keV energy band down to a 
limiting flux of $\sim 4\times 10^{-15}$ \ecs.
The (0.5 - 2.0) keV energy band was 
preferred over the broad band energy band (0.1-2.4 keV) in order to minimize 
any contribution from Galactic emission which can dominate below 0.5 keV.
Furthermore, due to the rapid increase in the size of the point spread 
function with off-axis angle, the detection analysis was limited to the 
inner 18 arcmin of the field centre in each PSPC image.
The total area surveyed at the $5 \sigma$ limit is therefore 1.41 deg$^2$.
The  X-ray flux limit depends on 
the exposure time, on the galactic N$_H$ in the direction of 
each field and on the position of the sources 
with respect to the centre of the field. 
The sky coverage and the procedure adopted to compute it are reported in 
Boyle et al., 1994. 
Conversion factors appropriate for a power-law spectrum with 
energy spectral index of $\alpha_x = 1$ and for the galactic N$_H$ in the 
direction of each field were used for the conversion between
counts and flux (see Georgantopoulus et al., 1995 for details).
This conversion factor is accurate to $\pm 10\%$ for all the energy 
spectral indices in the range $0.5 < \alpha_x < 1.5$ for the low 
Galactic $N_H$ values along the line of sight of the used fields. 

X-ray astrometric corrections were applied by using the optical
positions of the known 10-12 Durham UVX AGN detected by ROSAT in each
field. The search was made for the optical counterparts of the other
X-ray sources detected out to a radius of $\sim 20^{\prime\prime}$ of the
trasformed X-ray position.  Optical spectroscopy for these counterparts
was performed by using the automated fibre-optic system (AUTOFIB) and
the Low-Dispersion Survey Spectrograph (LDSS) at the Anglo-Australian
Telescope (AAT).  Amongst the sources so far identified, there are 107
BLAGN, 7 early-type galaxies, 12 NELG, 1 cluster
of galaxies and 12 stars (Georgantopoulos et al 1995).  Of the 49
sources with no optical identification, in 30 cases the spectra were too
poor to permit a reliable identification.

\bigskip
As previously noted by Stocke et al. (1983) and subsequent works, 
the X-ray to visual flux
ratio is a powerful tool in the process of optical identification of
X-ray sources.  We have identified $\sim 75\%$ of the 194 X-ray sources
found in our fields and are now in a position to study
the log($f_x/f_B$) distribution of the various source classes.

In figure 1 we show the X-ray to optical flux ratios for all classes of
X-ray sources found in our ROSAT deep surveys: panel (a) - QSOs; panel
(b) - NELGs, Early-type Galaxies and Stars.  The
relation $$ log[f_x(0.5-2.0 kev)/f_B] = log(f_x(0.5-2.0 kev)) + B_J/2.5
+ 5.5$$ has been adapted from the original relation $$ log[f_x(0.3-3.5
kev)/f_V] = log(f_x(0.3-3.5 kev)) + V/2.5 + 5.37$$ given in Maccacaro et
al. (1988) assuming a power-law X-ray spectral model with energy index
$\alpha_x = 1.0$ and an optical colour-index (B--V) =0.3.  The two dotted lines in figure 1
indicate the range of log($f_x/f_B$) comprising $\sim 70\%$ of the
sources identified with BLAGNs in the EMSS, while the two dashed lines
indicate the range of log($f_x/f_B$) comprising $\sim 70\%$ of the
sources identified with Galaxies.  The X-ray to optical flux ratio of
our sample of BLAGN is in very good agreement with the same ratio for the
BLAGN in the EMSS sample, i.e. the two dotted lines include $\sim 70\%$ of
the BLAGN in our sample.  Since our BLAGN sample has a higher mean redshift
($<z> \sim 1.5$) than those in the EMSS sample ($<z> \sim 0.4$), this
figure clearly shows that the X-ray to optical properties of X-ray
selected BLAGN do not change very much with $z$.  This is consistent with
the results obtained for optically selected BLAGN (Avni and Tananbaum, 1986;
Wilkes et al., 1994).  The NELGs
and early-type galaxies from our sample occupy a part of the
$f_X/f_B$ diagram located between those parts occupied by BLAGN and normal
galaxies, although there are 4 NELGs (three belonging to the
``restricted'' ROSAT sample defined below) which lie
within the AGN locus and which are therefore suspected of AGN activity.
The X-ray and optical properties of the 12 NELGs belonging 
to our sample are reported in Table 1; the optical spectra are presented 
in figure 2. A more detailed discussion of their optical properties
is presented in Griffiths et al. (1995).

\bigskip
In order to reduce the problem of field contamination, we have selected
and used in the following analysis a subsample of NELGs (hereafter the
``restricted'' sample), defined to have an offset between the X-ray and
optical positions less than $20^{\prime\prime}$, together with a
magnitude limit of B$_J < 21.5$.  This ``restricted'' sample consists of
the 7 NELGs annotated with a ``C'' in column 13 of Table 1.

We have addressed the problem of the contamination from the field
galaxies in this ``restricted'' sample in the following way.
In the error circles of the X-ray sources with spectroscopically
identified QSOs, we have recorded the angular distance from the X-ray
centroid to the nearest non-stellar objects having apparent magnitude
less than B.  The distribution of the recorded objects, as a function of
the distance, has then been normalized to the 75 X-ray source fields
which do not contain identified QSOs or stars, and this distribution has
been compared with the similar distribution for the galaxies
spectroscopically identified in the survey.  We have thus estimated that
about 2 sources with $B < 21.5$ are expected by chance to be the
nearest neighbours of these 75 X-ray sources, at distances less than 20
arcsec.  
If we bear in mind that galaxies with emission lines are about
$20\%$ of the field galaxy population, we expect less than one object
to be a chance galaxy in our ``restricted'' sample of NL objects.

\subsection{Previous Samples of X-ray Selected Narrow Line Galaxies }\tx
\noindent

The CRSS (Boyle et al.,1995a) is a study of serendipitous X-ray sources in 
20 PSPC fields at
high galactic latitude ($|b^{II}| > 30^o$).  In order to simplify the
optical identification process, the X-ray source detection in each field
was limited to a flux level $f_x$ (0.5 - 2 keV) $> \ 2\times 10^{-14}$
\ecs .  About $90 \%$ of the 123 x-ray sources in the sample were
spectroscopically identified and amongst them there were 10 NELGs 
(we have excluded from the original sample of 12 NELGs 
reported in Boyle et al. (1995a) the two objects (CRSS1413.3+4405 and 
CRSS1415.0+4402) now classified as Seyfert 1.5 in Boyle et al. (1995b).
High resolution, high signal-to-noise optical
spectroscopy allowed these authors to classify the NELGs
as starburst-like or Seyfert 2 galaxies, with approximately equal numbers.
This sample of NELGs is important for our purposes because it
fills the gap between those found in our ROSAT deep surveys
($f_x$ (0.5 - 2 keV) \gaeq $3\times 10^{-15}$ \ecs ) and those
in the EMSS ($f_x$ (0.3 - 3.5 keV) \gaeq $1\times 10^{-13}$ \ecs ).

The {\it Einstein Observatory} EMSS (Gioia et al., 1990; Stocke et al.,
1991; Maccacaro et al., 1994) contains several examples of NELGs (having
emission lines with FWHM $<$ 1000 km s$^{-1}$).  As shown by Fruscione
and Griffiths (1991) and Fruscione, Griffiths and Mackenty (1993) these
objects can generally be found amongst the EMSS X-ray sources associated
with galaxies with ``ambiguous'' classification, with spectral
properties which were either unclear or at the borderline between those
of AGN and normal/starburst galaxies.  The ambiguous EMSS X-ray sources
which satisfy the above requirements are listed in Tables 8 and 10 of
Stocke et al. (1991), numbering 36 objects in total.  However, in many
cases the entries in these tables are simply due to insufficent signal
to noise ratio at $H_{\alpha}$ or $H_{\beta}$ to determine unambiguously
if a broad-line component is present. As explicitly reported by Stocke
et al., higher resolution and higher signal-to-noise spectra
would clarify whether weak broad-line components are present in many of
these objects.

In an attempt to investigate the presence of ``starburst'' galaxies in
the EMSS sample we have taken optical spectra of 15 EMSS NELG
 candidates and have used the ratios of selected emission lines to
classify the observed galaxies.  The observations were made during 3
nights in February and October 1991 at the Steward Observatory 2.3m
telescope, equipped with a Boller and Chivens spectrograph, and 2 nights
in October 1991 at the Multi Mirror Telescope Observatory 4.45m
equivalent telescope plus ``Red'' spectrograph. The detector was in both
cases a Texas Instruments 800 $\times$ 800 pixel CCD.  We took $\sim$10
\AA--resolution spectra and oriented the slit along the major axis of
the galaxies covering a total spectral range of $\approx 4500-8200$
{\AA}. We reduced the data using standard IRAF packages and measured the
flux and the width of selected emission lines from the extracted 1-d
spectra using the task "splot". We used the option of fitting a single
Gaussian profile or else deblending multiple Gaussians (e.g. in the case of
H$\alpha$ and [NII]). The flux and FWHM are calculated by this task for the
fitted Gaussian profiles above the continuum level.  We estimated an
average 20-30\% uncertainty in the flux level.  We computed the redshift
of each source by measuring the centroid of a gaussian centered
on the [O~III]$\lambda$5007 or H$\alpha$ emission line. In all cases the
redshift corresponds to the redshift reported in Stocke et al.,
confirming the identification.

The spectra of the 15 observed EMSS objects are shown in figure 3.  We
considered both the broadening of the line profiles and the line-ratio
diagnostics in the classification of the objects. Lines with FWHM \gaeq
1000 km sec${^-1}$ were considered an indication of a broad line region,
i.e. indicative of a BLAGN. The line-ratio
diagnostics described in Veilleux and Osterbrock (1987) were used to
discriminate between HII and Sy2 types: however, in some cases, the
comparision of different line-ratio gives a different classification, or
the position in the dignostic diagrams is ambigous (see e.g. Fig. 3 in
Fruscione and Griffiths 1991). In these cases the objects were
classified as borderline HII/Sy2.  The emission line properties and
classification for the observed objects, along with the emisison lne
properties and classification for other ``ambiguous'' EMSS sources from
the literature, are reported in Table 2.  The (0.3-3.5) keV X-ray
fluxes, redshifts and V magnitudes reported in Table 2 were taken from
Stocke et al. (1991).  X-ray fluxes were corrected for galactic
absorption and were computed assuming a power law spectral model with
energy index $\alpha_x =1$.

\subsection{Overall Properties of the ROSAT/EMSS sample}\tx
\noindent

The spectroscopic properties of the NL objects in the EMSS sample are
similar to those of the NELGs that we have identified here (see 
Griffiths et al., 1995), as well as
those in the CRSS.  For the following analysis, we need to define a
working sample of X-ray selected NELGs.  We have thus taken the 7
objects from the ``restricted'' ROSAT sample of NELGs (sources
annotated with a ``C'' in Table 1), the 10 NELGs in the CRSS sample and
the 15 NELGs in the EMSS sample (sources annotated with a ``C'' in Table
2).  Amongst the 10 NELGs in the CRSS sample there are 5 HII region-like
galaxies, 2 Seyfert 2's, 1 Seyfert 1.8 and two objects with ambiguous
classification.  Amongst the 15 in the EMSS sample there are 6 HII
region-like galaxies, 5 Seyfert 2's, 2 LINERs and 2 objects at the
borderline between HII and Seyfert 2 types.  We have added the two
LINERS to the NELG sample because recent results on optical
spectroscopy of luminous infrared galaxies (Veilleux et al., 1995),
show that circumnuclear starburst activity is a common feature 
for this class of objects.

The sky coverage utilized in the present paper is a combination of the
EMSS sky coverage (reported in Gioia et al.(1990)), the ``ROSAT
effective survey area'' of our fields (reported in Table 2 of Boyle et
 al., 1994) and the ``ROSAT effective survey area'' of the CRSS survey
(see Boyle et al., 1995a).  The ``ROSAT effective survey area'' takes
into account the spectroscopic incompletness in the process of
identification of the faint sources.  The three sky coverages as a
function of X-ray flux have been computed assuming a power law
spectrum with energy index $\alpha_x = 1$.  The (0.5 - 2.0) keV X-ray
fluxes have been converted into the (0.3 - 3.5) keV energy band (the
EMSS energy band) assuming the same spectral index.

In figure 4 we show the distribution of the sample of NELGs in the
$L_x-z$ plane.
The ROSAT NELGs (both our sample and the CRSS sample) have X-ray
luminosities in excess to $\sim 10^{42}$ \es and lie in the redshift
range $\sim 0.1\div 0.6$.  Similarly, 12 of the EMSS NELGs have an
X-ray luminosity in excess of $\sim 10^{42}$ \es and the 2 highest
redshift EMSS starburst galaxies have a luminosity in excess of $\sim
10^{43}$ \es.
The distribution of NELGs in the $L_x-L_B$ plane is shown in
figure 5.
$L_x$ is the X-ray luminosity in the (0.3-3.5 keV) energy band and $L_B$
is the optical B luminosity in $L_{\odot}$ (see Canizares, Fabbiano and
Trinchieri, 1987 for the definition of $L_B$).  For the EMSS objects,
B magnitudes were computed from the original V magnitudes reported in
table 2, assuming B--V = 1.1, 0.8 and 0.3 according to the prescription
given in Stocke et al.(1991) (see also the notes to Table 2).
Palomar O magnitudes for 8 of the objects in the CRSS sample are reported 
in Boyle et al., 1995b. For the  remaining 
two objects, CRSS1514.4+5627 and CRSS1605.9+2554, 
we have only Palomar E magnitudes and they are 20.69 and 20.55, respectively.
For the CRSS objects we have used the Palomar O or Palomar E magnitudes.
The shaded line in figure 2 encloses the region populated by the normal
late-type galaxies observed with the {\it Einstein} Observatory
(Fabbiano, Kim and Trinchieri, 1992).  The optical luminosity of our
ROSAT sample ranges between $L_B \sim 10^{10} L_{\odot}$
($M_B \sim -20$) to $L_B \sim 10^{11} L_{\odot}$ ($M_B \sim -22$); this
range of optical luminosities is typical of large spiral or elliptical
galaxies.  However, for a given optical luminosity, the X-ray luminosity
of the NELGs is about one to two orders of magnitude
higher than the X-ray luminosity of normal late-type galaxies,
suggesting a different and/or more efficient X-ray emission mechanism.

\section{The Log N($>$S)-Log S of NELGs}\tx
\noindent

In figure 6 we show the Log N($>$S)-Log S relationship of our working
sample of NELGs (filled circles).  This Log N($>$S)-Log S has been
obtained by folding the total sky coverage with the flux of each source.
It is now interesting to compare this Log N($>$S)-Log S with the BLAGN Log
N($>$S)-Log S obtained from the combined EMSS-ROSAT sample.  The BLAGN Log
N($>$S)-Log S is shown in figure 6 as open triangles. 
There is an apparent increase of the surface density of NELGs
with regard to the surface density of BLAGN towards lower fluxes.
The ratio between NELGs and BLAGN increases from $\sim 0.04$ at $f_x$
\gaeq $6\times 10^{-13}$ \ecs to $\sim 0.1$ at $f_x$ \gaeq $10^{-14}$ \ecs.
If this trend is confirmed by better statistics and deeper X-ray surveys
(e.g. the AXAF and XMM deep surveys) then the spatial density of 
NELGs could be very close to the spatial density of BLAGN at $f_x \sim
10^{-15}$ \ecs and could produce a re-steepening of the total Log
N($>$S)-Log S for $f_x < 10^{-15}$ \ecs.  It is interesting to note that
such re-steepening is not inconsistent with the P(D) fluctuations
analysis of the deepest ROSAT fields (Georgantopoulos et al. (1993), 
Hasinger et al. (1993), Barcons et al. (1995)).

We may also compare the Log N($>$S)-Log S of X-ray selected NELGs
with the predictions of Griffiths and Padovani (1990), which were based
on an infrared selected (IRAS) sample of starburst galaxies and on the
observed relation between infrared and X-ray luminosities.  The
prediction shown in figure 6 (dotted line) refers to the case of a
cosmological luminosity evolution given by $L_x(z) \propto e^{C\tau}$
where $\tau$ is the look-back time and C is the cosmological
evolution parameter (C=5 for this particular curve).  On the assumption
that about half of the NELGs are starburst in origin, we found a
factor $\sim 6$ more galaxies with respect to the prediction, suggesting
that these objects have higher X-ray luminosities than those predicted
by GP90, which were based on the $L_x/L_{IR}$ ratio.

A number of models, based on the unification schemes of AGNs, have been
recently proposed to explain the origin of the diffuse cosmic X-ray
Background.  In particular, Comastri et al. (1995) and Madau, Ghisellini
and Fabian (1994) have shown that a mixture of unabsorbed AGN (Seyfert 1
galaxies and QSOs) and absorbed AGN (Seyfert 2 galaxies), integrated
over the luminosity-redshift plane can account for the cosmic X-ray
background from several to $\sim 100$ keV.  It is thus interesing to
compare the predicted surface density of the hypothetical absorbed
population with our observed Log N($>$S)-Log S.  The models of Comastri
et al. and Madau, Ghisellini and Fabian predict a surface density of
absorbed type 2 AGNs ($N_H > 10^{22}$ cm$^{-2}$) of $\sim 25$ deg$^{-2}$
at $f_x (0.3 -3.5 keV) \sim 10^{-14}$ \es.  This prediction is a factor
2 above our observed Log N($>$S)-Log S of the total population of NELGs,
including Seyfert 2 and starburst galaxies. Owing to the incomplete
spectroscopic identification and the small numbers, it may be premature
to say if this discrepancy is statistically significant.
 
Some incompleteness at fluxes fainter than $3\times 10^{-14}$ \ecs is
also suggested from the shape of the number--counts relationship.  The
CRSS sample and the EMSS sample have a high rate of optical
identification ($\geq 90\%$), while in our sample the identification
rate is of the order of $75\%$.  We have therefore corrected, to first
order, for the incompleteness following Boyle et al., 1993, but a higher
fraction of NL galaxies is probably present amongst the unidentified
sources (which are, in the mean, at fainter fluxes) when compared with
the identified NL objects as a fraction of the total identified sources.
We will discuss this problem and its influence on the cosmological
evolution analysis in the next section.

\section{The Cosmological Evolution and X-ray Luminosity Function of
NELGs}\tx

The $V_e/V_a$ test (Avni and Bahcall, 1980) is a powerful tool for
studying the cosmological evolution of any class of objects.  If the
test is applied to a sample that is known to be complete, then the
quantity $V_e/V_a$ has the property of being uniformly distributed
between 0 and 1, with a mean value of 0.5 in the absence of
cosmological evolution.
For our sample of 32 NELGs we find that the hypothesis of a uniform
distribution in space is rejected at more than the $99.99\%$ confidence
level ($<V_e/V_a> = 0.73 \pm 0.05$).

The evolution can be detected up to a certain redshift limit, where 
the evaluation of this limit depends on being able to
sample the observed luminosity function within different redshift intervals 
(see e.g. Maccacaro et al., 1991; Boyle et al., 1993 for results 
on the evolution of the X-ray luminosity function of BLAGN).
With a total number of $\sim 30$ objects we cannot follow this 
approach. 
We have thus applied the $V_e/V_a$ test (see below) over the redshift 
range 0.0 -- 3.0, but the results are essentially unchanged if we restrict 
the analysis to the redshift range 0.0 -- 2.0. 
In other words, with the data at our disposal, we cannot say if the 
evolution of these objects ``switches off" at $z\sim 3$ or at $z \sim 2$ 
(as found for optically selected QSOs and X-ray selected BLAGN by 
Boyle et al., 1991 and Boyle et al., 1993) or at a lower redshift.
Better statistics are needed to measure this behaviour,
such as those anticipated from the AXAF and/or the XMM deep surveys.

For comparison with other classes of active extragalactic objects,
we note that Rowan-Robinson et al. (1993) have tested a range of different
evolutionary models (either luminosity evolution and/or density
evolution), and have shown that the faint radio source counts can
be explained by a population of starburst galaxies undergoing strong
luminosity evolution characterised by $L_x(z) \propto (1+z)^C$.  The
cosmological evolution rate they found ($C\sim 3.1$ in a q0=0.5
Friedmann universe) is very close to that found for radio galaxies and
quasars (Dunlop and Peacock, 1990), optically selected QSOs (Boyle et
al.(1990)) and X-ray selected AGN (Boyle et al., 1994).  Furthermore, as
shown by Boyle et al. (1990) for optically selected AGN and by Della
Ceca et al. (1992) and Boyle et al. (1994) for X-ray selected AGN, a
cosmological evolution law characterised by $L_x(z) \propto (1+z)^C$
provides a better description of the current BLAGN data set.

In order to compare directly our results with those of other authors, we
assume a pure luminosity evolution model with the evolutionary form
$$L_x(z) = L_x(0) \times (1+z)^C$$ where $L_x(0)$ is the de-evolved
(z=0) X-ray luminosity, $L_x(z)$ is the luminosity at redshift $z$ and C
is the cosmological evolution parameter.  The best fit C can be
determined by finding the value that makes $<V_e/V_a> = 0.5$ and the
individual $V_e/V_a$ values uniformly distributed between 0 and 1.  The
$1\sigma$ interval on C corresponds to the values for which $<V_e/V_a> =
0.5 \pm (12N)^{-1/2}$, where N is the number of objects in the sample.
The best fit value we have found for the evolution parameter is C=3.35
with associated $1\sigma$ and $2\sigma$ confidence intervals of
[3.09-3.57] and [2.72-3.79] respectively.

In order to evaluate how this result is affected by the uncertainty on
the exact number of NELGs in the EMSS sample, by the 
contamination problem in {\it our} sample and by the unidentified
objects in ours and the CRSS sample we have considered three extreme
cases.  First of all, we have considered {\it all} the objects listed in
Table 8 and 10 of Stocke et al. (1991), for which we have no evidence of
a broad line component, as possible EMSS NELGs.  This sample is then
comprised of 24 EMSS objects (we have added to the original EMSS NELG
sample the 9 objects annotated with an ``E'' in table 2) and can be
considered as a useful upper limit to the real number of NELGs belonging
to the EMSS.  The cosmological evolution analysis of these 24 EMSS
objects taken together with the 17 ROSAT NELGs (10 CRSS objects plus 7
from our ROSAT deep surveys) gives $<V_e/V_a> = 0.71 \pm 0.05$, still
indicating cosmological evolution at the $99.99\%$ confidence level.
The best fit evolution parameter is C=3.15 with associated $1\sigma$ and
$2\sigma$ confidence intervals of [2.88-3.35] and [2.50-3.54]
respectively.

Secondly, in order to evaluate the effect of the contamination problem,
we have excluded from the original working sample of 32 NELGs the
two NELGs in our ROSAT sample with optical magnitude greater than
21 (GSGP4X:91 and GSF1X:36). These two objects have the highest
redshifts in our sample (see Table 1).  Analysis of the sample comprised
of the remaining 30 objects gives $<V_e/V_a> = 0.71 \pm 0.05$, still
indicating cosmological evolution at the $99.99\%$ confidence level.
The best fit cosmological evolution parameter is, in this case, C=3.36
with associated $1\sigma$ and $2\sigma$ confidence intervals of
[3.06-3.60] and [2.58-3.85] respectively.

Thirdly, we have considered the total sky coverage for our sample
(reported in Table 2 of Boyle et al., 1994) and for the CRSS sample
(Boyle et al., 1995a), on the assumption that no more NELGs were to be
found among the unidentified objects.  Analysis on our working sample of
32 NELGs then gives $<V_e/V_a> =0.72 \pm 0.05$, still indicating
cosmological evolution at the $99.99\%$ confidence level.  The best fit
evolution parameter is, in this case, C=3.16 with associated $1\sigma$
and $2\sigma$ confidence intervals of [2.86-3.41] and [2.42-3.62]
respectively.  However, it may realistically be assumed that more NELGs
will be found amongst the unidentified ROSAT sources.
This is also suggested from the shape of the number-counts relationship
for sources with fluxes f(x) \laeq $4\times10^{-14}$\ecs  Since these
objects are, in the mean, at the faintest X-ray fluxes their $V/V_{max}$
will be higher than the mean $V/V_{max}$ of the present sample.  As a
consequence, the evolution parameter C is expected to be (slightly)
higher than the value derived here.

These results show that, although the statistics are poor (32 objects in
total), we have evidence of cosmological evolution of NELGs at a high
confidence level.

We briefly mention the different physical scenarios which could be
responsible for the evolution of the X-ray properties of NELGs.  Under
the hypothesis that their X-ray emission derives primarily from massive
X-ray binaries (MXRB) Griffiths and Padovani (1990) have pointed out
that there may be an inverse correlation between their luminosity and
the metallicity of the host galaxy.  Since galaxies at high redshift may
be poorer in metals than the present-day ones, the implication would be
that their X-ray luminosity would be correspondingly higher.
Furthermore, the star formation and supernova rates would be higher than
those in local galaxies.  Starbursting galaxies at moderate to high
redshift may thus contain a larger number of MXRBs than our Galaxy. On
the other hand, we may be observing an evolving AGN component in these
objects. In the latter case, we should perhaps expect the evolution 
to proceed at a similar rate to that of {\it bona fide} or established AGN.

Using the derived best fit value for the cosmological evolution
parameter (C = 3.35) we can now estimate a de-evolved (z=0) XLF of NELGs
starting from an X-ray selected sample.  A non-parametric
representation of the de-evolved XLF is obtained using the Avni and
Bahcall (1980) estimator $\sum 1/V_a$ where $V_a$ is the volume
available within which a given object can be seen and still be part of
the sample. $V_a$ takes into account the effects of evolution (see
Maccacaro et al. (1991) and Della Ceca et al. (1992) for more details).
The resulting X-ray luminosity function of the NELGs is shown in
figure 7 (filled circles).  The data have been binned in equal
logarithmic widths of 0.5 and $1\sigma$ error bars have been computed
using Poisson statistics (Gehrels, 1986).  Obviously, the de-evolved
luminosity function shown in figure 7 is model-dependent and, in
general, steeper evolutionary laws and higher values of C lead to
steeper luminosity functions.  Due to the poor statistics of our sample
the cosmological evolution parameter C is not well constrained.  To show
how this uncertainty has repercussions on the de-evolved XLF we have
also reported in figure 7 (dotted lines) the de-evolved XLFs obtained
using the $90 \%$ confidence interval for the cosmological evolution
parameter C: the de-evolved XLF we have derived is not a strong function
of C.

We have also performed a maximum likelihod analysis to obtain a ``best fitting''
parametric representation of the evolution and luminosity function 
(see Boyle et al., 1994 for the application of this method).
The local (z=0) X-ray luminosity function (XLF) has been described by a 
power law with two components:

$$\Phi_x(L_x) = \cases{K_1 L_{x_{44}}^{-\gamma_1}, & $L_x (z=0) < L_x^*$; \cr
                       K_2 L_{x_{44}}^{-\gamma_2}, & $L_x (z=0) > L_x^*$.\cr} $$

\noindent where $K_1$, $\gamma_1$ and $\gamma_2$ represent the
normalization, and the faint and bright end slopes of the XLF, respectively. 
The quantity $L_{x_{44}}$ is the luminosity in the (0.3--3.5) keV energy
band expressed in units of $10^{44}$ erg s$^{-1}$.  The faint-end and
bright-end normalizations, $ K_1 $ and $ K_2 $, are tied together by the
requirement of continuity of the XLF at $L_{x_{44}}^*$, which implies
$K_2 = K_1 / L_{x_{44}}^{*(\gamma_1 - \gamma_2)}$.

Due to the small number of objects at our disposal and their limited
redshift range, the parameter C was not well constrained when we tried
to fit simultaneously the evolution and the shape of the XLF.
We then fixed it at the value obtained using the $V/V_{max}$ analysis 
(C=3.35) and determined the best fit values for the shape of the XLF.
We obtain the following `best-fit' values: $\gamma_1 = 1.85\pm 0.25$,
$\gamma_2 = 3.83\pm 0.20$, $L_x^* = 10^{42.83\pm 0.2}$ and $K_1 =
1.3\times 10^{-7}\ Mpc^{-3} (10^{44} erg s^{-1})^{-1}$.
This model was an acceptable fit to the data with a KS probability of
greater than 20 per cent, according to the 2D KS-statistic (see Boyle et al.
1994 for details of this test).
  The derived $z=0$ ``best fit'' XLF is
reported in figure 7 as a solid line.  The dashed line in figure 6 shows
the predicted number-flux relationship for the NELGs, obtained by
integrating our best fit model over the luminosity range $L_x = 10^{40} -
10^{45}$ \es and out to a redshift of 3.

We now compare the de-evolved XLF of the NELG with the de-evolved XLF of BLAGN 
determined by Boyle et
al. (1994) from the combined EMSS - ROSAT sample.  The BLAGN XLF (model K
of Boyle et al., 1994) is shown in figure 7 (dashed line).  The XLF of
NELGs in the luminosity range $10^{41} - 3\times 10^{43}$ \es is
steeper than the BLAGN XLF in the same luminosity interval.  The spatial
density of NELGs is significantly lower than the spatial density
of BLAGN at $L_x \sim 5\times 10^{43}$ \es; this difference decreases at
lower luminosities and at $L_x \le 10^{42}$ \es the spatial density of
NELGs is very close to that of BLAGN.

We have compared the spatial density of NELG with the spatial density of BLAGN 
at $z=0.0$. 
Given that the two kind of objects show similar cosmological evolution the 
density ratio NELG/BLAGN at $L_x(z)$ can be obtained using figure 7 and 
considering that $L_x(z=0) = L_x(z) \times (1+z)^{-C}$.

Having determined the cosmological evolution and the de-evolved
XLF of NELGs, starting from X-ray data, we are now in a position
to estimate {\it directly} their contribution to the cosmic X-ray background.
Using the formalism of Maccacaro et al. (1991) with our best fit 
parameter for the cosmological evolution (C = 3.35) and the estimated XLF at 
$z = 0$ (figure 7), we have been able to compute the percentage 
contribution to the 2 keV X-ray background in different bins of 
luminosities and redshifts.
We have used a 2 kev XRB intensity ($I_{XRB}$) equal to 
6.14 keV cm$^{-2}$ s${-1}$ sr${-1}$ keV$^{-1}$ 
(Gruber 1992), which is consistent within the 
calibration uncertainties with the values obtained by Hasinger (1992) from 
 ROSAT PSPC data ($I_{XRB}$ = 6.61  keV cm$^{-2}$ s${-1}$ sr${-1}$ keV$^{-1}$).
The results are reported in Table 3.
The principal uncertainty in this calculation is the faint 
end ($L_x$ \laeq $\sim 2 \times 10^{42}$ \es ) slope of the de-evolved XLF.
To take account of this uncertainty  we have reported in Table 3 
the percentage contribution to the 2 keV X-ray background for two 
extreme cases.  
The lower limit in Table 3 refers to a faint end slope $\gamma_1 = 1.6$ 
while the upper limit refers to a faint end slope $\gamma_1 = 2.1$.
These values refer to $\pm 1 \sigma$ error on $\gamma_1$.
NELGs can thus account for 
$\sim 11-36 \%$ of the 2 keV X-ray background. 
The results we have obtained in this analysis thus confirm and strengthen 
the results obtained by Boyle et al (1995a).
More than 55 $\%$ of the total NELG contribution at 2 keV comes from objects 
with $L_x > 10^{41}$ \es, i.e. from objects that we have already seen.
About two thirds of the overall contribution comes from objects with 
$z < 2$.

We have also checked the results reported in Table 3 as a function of
the cosmological evolution parameter(s) used in the calculations.  Using
values of 2.72 and 3.79 for C ($95 \%$ confidence interval) and the
corresponding ``best fit'' XLFs, 
the fraction of the 2 keV XRB intensity accounted
for by the NELGs became $\sim 10 \%$ and $\sim 26 \%$, respectively.
The principal uncertainty in the contribution to the XRB is thus the
faint end slope of the XLF, rather than the current uncertainty in the 
evolution parameter.

\section{Summary and Conclusions}\tx
\noindent

We have used a sample of 7 X-ray emitting NELGs found in 5 deep ROSAT
fields together with similar samples from the CRSS (10 objects) and from
the {\it Einstein} EMSS (15 objects) to investigate and clarify their
cosmological properties and evolution.  The X-ray luminosities of NELGs range
from $10^{41} - 10^{44}$ \es and, for a given optical luminosity, their
X-ray luminosities are about one or two orders of magnitude greater than
those observed for typical late type galaxies.

Using these data we have been able to compare the number density of
NELGs and BLAGN as a function of flux. This ratio increases from $\sim
0.04$ at $f_x$ \gaeq $6\times 10^{-13}$ \ecs to $\sim 0.1$ at $f_x$ \gaeq
$10^{-14}$ \ecs, suggesting that the surface density of NELGs could be
very close to that of BLAGN at $f_x \sim 10^{-15}$ \ecs.  We find that
NELGs have a similar cosmological evolution rate to BLAGN, but NELGs have
a steeper XLF (in the interval $10^{41} - 3\times 10^{43}$ \es ).  Their
volume density is significantly lower than that of X-ray selected BLAGN at
$L_x \sim 5\times 10^{43}$ \es, but this difference decreases at lower
luminosities such that at $L_x \le 10^{42}$ \es the volume density of
NELG and BLAGN objects is very close.  Starting with an X-ray selected
sample of objects and having determined their XLF and cosmological
evolution, we have been able to directly estimate their contribution to
the 2 kev X-ray background.  Based on the observed range in the
parameter values for the best-fit evolutionary models, these objects can
account for $\sim 11 - 36 \%$ of the 2 kev X-ray background.

\section*{Acknowledgments}\tx

REG acknowledges receipt of NASA awards NAG5-1697, NAGW-1472 and
NAGW-3288. 
BJB acknowledges receipt of a Royal Society University
Research Fellowship.  We thank the staff of the Anglo-Australian
Telescope for their help and support in facilitating the optical
observations.
This research has made use of the NASA/IPAC Extragalactic Database 
(NED) which is operated by the Jet Propulsion Laboratory, Caltech, 
under contract with the National Aeronautics and Space Administration.

\section{References}\tx

\noindent\bibitem
Avni Y.,  Bahcall J.N., 1980, ApJ, 235, 694

\noindent\bibitem
Avni Y.,  Tananbaum H., 1986, ApJ, 305, 83

\noindent\bibitem
Barcons X., Franceschini A., De Zotti G., Danese L., Miyaji T., 1995, 
ApJ, in press

\noindent\bibitem
Barcons X.,  Fabian A.C., 1992,
``The X-ray Background'', Cambridge University Press, Cambridge

\noindent\bibitem 
Benn C.R., Rowan-Robinson M., McMahon R.G., Broadhurst T.J., 
Lawrence A., 1993, MNRAS, 263, 98 

\noindent\bibitem 
Boller Th., Meurs E.J.A., Brinkmann W., Fink H., Zimmermann U.,  
Adorf H.-M., 1992, A\&A, 261, 57.

\noindent\bibitem 
Boller, Th., Brandt, W. N. \& Fink, H., 1995, A\&A, in press

\noindent\bibitem
Boyle B.J., Fong R., Shanks T.,  Peterson B.A., 1990, 
MNRAS, 243, 1.

\noindent\bibitem
Boyle B.J., Griffiths R.E., Shanks T., Stewart G.C.,  Georgantopoulus,
 I., 1993, MNRAS, 260, 49.

\noindent\bibitem
Boyle B.J., Shanks T., Georgantopoulos I., Stewart G.C.,  
Griffiths R.E., 1994, MNRAS, 271, 639

\noindent\bibitem
Boyle B.J., McMahon R.G., Wilkes B.J., Elvis M., 1995a, MNRAS, 272, 462

\noindent\bibitem
Boyle B.J., McMahon R.G., Wilkes B.J., Elvis M., 1995b, MNRAS, in press

\noindent\bibitem 
Canizares C.R., Fabbiano G.,  Trinchieri G., 1987, ApJ, 312, 503.

\noindent\bibitem 
Comastri A., Setti G., Zamorani G., Hasinger G., 1995, A\&A, 296, 1

\noindent\bibitem 
Coziol, R., 1995, A\&A, in press

\noindent\bibitem 
David L.P., Jones C.,  Forman W., 1992, ApJ, 369, 121

\noindent\bibitem
de Zotti G., Toffolatti L., Franceschini A., Barcons X., Danese L.  \&
Burigana, C., 1995, in Bignami G., Ferrari A., Grindlay, J. Trumper,
J., Massaglia S., eds, Proc. International School of Space Science,
X-ray Astronomy

\noindent\bibitem
Della Ceca R., Maccacaro T., Gioia I.M., Wolter A.,  Stocke J.T., 
1992, ApJ, 389, 491

\noindent\bibitem 
Dunlop J.S.,  Peacock J.A., 1990, MNRAS, 247, 19

\noindent\bibitem 
Elvis et al. 1992

\noindent\bibitem 
Fabian A.C., 1988, in Kaiser N., Lasenby A. N., eds, The
Post-Recombination Universe, Kluwer Academic, p.51

\noindent\bibitem
Fabbiano G.F., 1989, ARAA, 27, 87

\noindent\bibitem 
Fabbiano G., Feigelson E., Zamorani G., 1982,  
ApJ, 256, 397

\noindent\bibitem 
Fabbiano G., Kim D.-W.,  Trinchieri G. 1992, 
ApJS, 80, 531

\noindent\bibitem 
Filippenko A.V., 1992, in Duschl W. J., Wagner S.J., eds,
Physics of Active Galactic Nuclei, Springer-Verlag, Berlin, p. 345 

\noindent\bibitem
Fruscione A., Griffiths R.E., 1991, ApJ, 330, L13

\noindent\bibitem
Fruscione A., Griffiths R.E., Mackenty J.W., 1993, in 
Observational Cosmology, ASP Conf.Ser., 51, 296

\noindent\bibitem
Gehrels N., 1986, ApJ, 303, 336

\noindent\bibitem 
Georgantopoulos I., Stewart G.C., Griffiths R.E., Shanks T., 
 Boyle B. J., 1993, MNRAS, 262, 619

\noindent\bibitem 
Georgantopoulos I., Stewart G.C., Griffiths R.E., Shanks T., 
 Boyle B.J., 1995, MNRAS, in press

\noindent\bibitem
Giacconi R., et al., 1979, ApJ, 234, L1

\noindent\bibitem
Gioia I.M., Maccacaro T., Schild R.E., Stocke J.T.,
Liebert, J.W., Danziger I.J., Kunth D., Lub J., 1984, ApJ, 283, 495 

\noindent\bibitem
Gioia I.M., Maccacaro T., Schild R.E., Wolter A., Stocke J.T.,
 Morris, S.L.,  Henry J.P., 1990, ApJS, 72, 567

\noindent\bibitem 
Green P.J., Anderson S.F.,  Ward M.J., 1992, MNRAS, 254, 30

\noindent\bibitem 
Griffiths R.E., Doxsey R.E., Johnston M.D., Schwartz D.A., 
Shwartz J.,  Blades J.C., 1979, ApJ, 230, L21

\noindent\bibitem 
Griffiths R.E., Murray S.S., Giacconi R., Bechtold J., Murdin P., Ward M.,
Peterson, B.A. Wright A.E.,  Malin D.F., 1983, ApJ, 269, 375

\noindent\bibitem 
Griffiths R.E,  Padovani, P., 1990, ApJ, 360, 483.

\noindent\bibitem
Griffiths R.E., Tuohy I.R., Brissenden R.J.V.,  Ward M.J., 1992, ApJ, 
255, 545.

\noindent\bibitem
Griffiths R.E., Georgantopoulus I., Boyle B.J., Stewart G.C.,  
 Shanks T., Della Ceca, R., 1995, MNRAS, 275, 77

\noindent\bibitem
Gruber D., 1992, in Barcons X.,  Fabian A.C., eds, 1992,
The X-ray Background, Cambridge University Press, Cambridge, p.44

\noindent\bibitem 
Hasinger G., 1992, in  Barcons X.,  Fabian A.C., eds, 1992,
The X-ray Background, Cambridge University Press, Cambridge, p. 229

\noindent\bibitem 
Hasinger G., Burg R., Giacconi R., Hartner G., Schmidt M., Truemper 
J.,  Zamorani G., 1993, A\&A, 275, 1

\noindent\bibitem 
Heckman T.M., 1991, in Leitherer C., Walborn N.R., Heckman T.M.,
Norman C.A., eds, Massive stars in Starbursts. Cambridge University
press, Cambridge, p. 289

\noindent\bibitem 
Henry J.P., Gioia I.M., Maccacaro T., Morris S.L., 
Stocke J.T., Wolter A., 1992, ApJ, 386, 408  

\noindent\bibitem 
Maccacaro T., Gioia I.M., Stocke J.T., 
1984, ApJ, 283, 486

\noindent\bibitem 
Maccacaro T., Gioia I.M., Wolter A., Zamorani G., Stocke J.T., 
1988, ApJ, 326, 680

\noindent\bibitem 
Maccacaro T., Della Ceca R., Gioia I.M., Morris S.L., Stocke J.T., 
 Wolter A., 1991, ApJ, 374, 117.

\noindent\bibitem 
Maccacaro T., Wolter A., McLean B., Gioia I.M., Stocke J.T., 
Della Ceca R., Burg R., Faccini R., 1994, Astro Lett\&Comm, 29, 267

\noindent\bibitem 
Madau P., Ghisellini G., Fabian A.C., 1994, MNRAS, 270, L17 

\noindent\bibitem 
Maiolino, R., Rutz, M., Rieke, G. H. \& Keller, L. D., 1995, 
ApJ, 446, 561

\noindent\bibitem 
Metcalfe N., Shanks T., Fong R., Roche N., 1995, MNRAS, 273, 257

\noindent\bibitem 
Moran E.C., Halpern J.P.,  Helfand D.J., 1994, ApJ, 433, L65

\noindent\bibitem
Mushotzky R.F., Done C., Pounds K.A., 1993, ARAA, 31, 717 

\noindent\bibitem
Norman C.A., Scoville N.Z., 1988, ApJ, 332, 124 

\noindent\bibitem 
Pfeffermann E. et al., 1986, Proc. SPIE, 733, 519.

\noindent\bibitem 
Primini F.A., Murray S.S., Huchra J., Schild R., Burg R.,  
Giacconi R., 1991, ApJ, 374, 440.

\noindent\bibitem 
Roche N., Shanks T., Georgantopoulos I., Stewart G.C., 
Boyle B.J., Griffiths R.E., 1995, MNRAS, 275, L15

\noindent\bibitem 
Rosati P., Della Ceca R., Burg R., Norman C., Giacconi R., 1995, ApJ, 
in press 

\noindent\bibitem
Rowan-Robinson M., Benn C.R., Lawrence A., McMahon R.G.,  Broadhurst,
 T.J. 1993, MNRAS, 263, 123

\noindent\bibitem 
Shanks T., Georgantopoulos I., Stewart G.C., Pounds K.A., Boyle B.J., 
 Griffiths R.E. 1991, Nat, 353, 315.

\noindent\bibitem 
Schnopper H.W., Davis M., Delvaille P., Geller M.J.,  Huchra J.P., 
1978, Nat, 275, 719

\noindent\bibitem
Stewart G.C., Fabian A.C., Terlevich R.J., Hazard C., 1982, MNRAS,
200, 61

\noindent\bibitem
Stocke J.T., Liebert J.W., Gioia I.M., Griffiths R.E., Maccacaro T.,
Danziger I.J., Kunth D., Lub J., 1983, ApJ, 273, 458

\noindent\bibitem
Stocke J.T., Morris S.L., Gioia I.M., Maccacaro T., Schild R., 
Wolter A., Fleming T.A.,  Henry J.P., 1991, ApJS, 
76, 813

\noindent\bibitem
Tresse L., Rola C., Hammer F., Stasinska G., in 
Maddox S.,  Aragon-Salamanca A., eds, Proc. 35th 
Herstmonceux Conf., Wide Field Spectroscopy and the Distant Universe,
World Scientific, Singapore

\noindent\bibitem 
Treyer M.A., Mouchet, M., Blanchard, A. \& Silk J., 1992, A\&A, 264, 11   

\noindent\bibitem 
Treyer M.A., Silk J., 1993, ApJ, 408, l1   

\noindent\bibitem
Truemper J., 1983, ASR, 2, 241

\noindent\bibitem
Veilleux S.,  Osterbrock D.E., 1987, ApJS, 63, 295 

\noindent\bibitem
Veilleux, S., Kim, D.-C., Sanders, D.B., Mazzarella, J.M., and
Soifer, B.T. 1995, ApJS, 98, 171

\noindent\bibitem 
Weedman D.W. 1986, in Carol J.  Lonsdale, ed, Star Formation in
Galaxies, NASA CP-2466, p.351.

\noindent\bibitem 
Wilkes B., Tananbaum H., Worrall D.M., Avni Y., Oey M.S., Flanagan J.,
 1994, ApJS, 92, 53  
 
\noindent\bibitem 
Williams R.J.R., Perry J.J., 1994, MNRAS, 269, 538

\noindent\bibitem 
Wilson, A. S., Elvis, M. S., Lawrence, A. \& Bland-Hawthorne, J., 1992,
ApJ, 391, L75

\section {Figure Captions.}\tx

\noindent {\bf Figure 1.} The X-ray to optical flux ratio for all
classes of X-ray sources found in the ROSAT deep surveys.  The two
dotted lines in figure 1 indicate the range of log[$f_x/f_B$] comprising
$\sim 70\%$ of the EMSS BLAGNs, while the two dashed lines indicate the
range of log[$f_x/f_B$] comprising $\sim 70\%$ of the EMSS normal
Galaxies.  Panel (a) : QSOs; Panel (b) : Narrow Emission Line Galaxies,
Early-type Galaxies and Stars.

\noindent {\bf Figure 2.} Optical spectra of the 12 Narrow Emission Line 
Galaxies found in our ROSAT fields. See Table 1 for more details. 

\noindent {\bf Figure 3.} Optical spectra of the 15 observed 
EMSS objects. See Table 2 for more details. The objects for which spectra
are shown are as follows: 

 MS0038.0+3242  MS0444.9$-$1000  MS1058.8+1003

 MS0038.7+3251  MS0942.8+0950   MS1110.3+2210 

 MS0116.7+0802  MS1019.0+4836   MS2044.1+7532

 MS0340.3+0455  MS1043.9+1400  MS2118.4$-$1050 

 MS0423.8$-$1247   MS1047.3+3518  MS2338.9$-$1206 

\noindent {\bf Figure 4} 
The X-ray luminosity - redshift distribution for 
our working sample of NELGs. The symbols are as follows: 
filled circles: our ROSAT NL sample; 
filled squares: HII region-like galaxies in the CRSS;
open   squares: Seyfert 2 like galaxies in the CRSS;
filled triangles : HII region-like galaxies in the EMSS;
open triangles   : Seyfert 2 like galaxies in the EMSS;
crosses : LINER in the EMSS;
open circles : ambiguous source in the EMSS and in the CRSS samples

\noindent {\bf Figure 5.} 
The distribution of the used sample of NELGs  in the
$L_x - L_B$ plane.  The shaded lines enclose the region populated by the
normal late-type galaxies observed with the {\it Einstein} Observatory
(Fabbiano, Kim and Trinchieri, 1992). See text for details.  The symbols
are as in figure 4.

\noindent {\bf Figure 6.}  The X-ray Log N($>S$) - Log S of the combined
(ROSAT + EMSS) sample of NELGs (filled circles) compared with the
BLAGN X-ray Log N($>S$) - Log S (open triangles).  The dotted line shows
the prediction of the X-ray Log N($>S$) - Log S of star-forming galaxies
based on a sample of infrared selected starburst galaxies.
The solid lines represent the EMSS BLAGN Log N($>S$) - Log S (best fit 
and $\pm 1 \sigma$ errors on the slope; from Della Ceca et al., 1992).
The dashed line represents the prediction on the number-flux relation 
based on the ``best fit" evolution model. See text for details.

\noindent {\bf Figure 7.}  The filled dots represent the differential
X-ray luminosity function of NELGs, computed at z=0 according to
the evolutionary law $$L_x(z) = L_x(0) \times (1+z)^C$$ with C=3.35
(best fit value).  Data have been binned with equal logarithmic widths
of 0.5; $1\sigma$ error bars are determined by the number of objects
which contribute to each bin and have been computed using Poisson
statistics.  The dotted lines represent the XLFs ($z = 0$) computed
using the $95 \%$ confidence interval on the cosmological evolution
parameter (C=2.72 and C=3.79).  The shaded lines represent the model K
of Boyle et al. (1994) for the X-ray luminosity function of X-ray
selected BLAGN.
The solid line represent the ``best fit" two-power law parametric 
representation of the XLF obtained using the maximum likelihood analysis.
See text for details.

\vfill\eject

\null

\vfill\eject




\font\sevenrm=cmr7    \font\fiverm=cmr5
\magnification=\magstep0
\newdimen\digitwidth
\setbox0=\hbox{\rm0}
\digitwidth=\wd0
\catcode`?\active       
\def?{\kern\digitwidth}
\hsize=19.5 true cm
\hoffset=-2.0 true cm
\voffset=0.0 true cm
\vsize=22 true cm
\raggedbottom
\baselineskip 10pt

\table{2}{D}
\noindent
\null
{\centerline {\bf Table 1. \rm Narrow Emission Line  Galaxies in our deep surveys.}}
\smallskip
\sevenrm
Column 1 gives the X-ray source name, columns 2 and 3 the X-ray field
position in degrees measured from the field centre, column 4 the radial
distance of the X-ray source from centre in arcmins, column 5 the X-ray
flux (\ergcms) (0.5 - 2.0 keV), columns 6 and 7 the celestial coordinates of
the X-ray source (equinox 1950.0), columns 8 and 9 the coordinates of
the optical counterpart (equinox 1950.0), column 10 the distance between
the X-ray source and optical counterpart in arcs, column 11 the B mag.,
column 12 the (U--B) colour, column 13 indicates with a ``C'' those
sources belonging to the ``Restricted'' ROSAT sample of NELGs (see
Section 2.1 for details), column 14 lists the redshifts, and column 15
contains comments.  In column 10, ``*'' indicates that the brighter
optical counterpart is listed and that there is also a fainter candidate
within the error circle.
In the
Comment column, W is the equivalent width of the dominant or single
narrow line, D is a measure of the ``H\&K'' break (the ratio of
continuum strength immediately above and below the CaII H \& K features
near 4000\A).

{
\settabs \+ xxxxxxxxxxx & xxxxxxx & xxxxxxx & xxxxxxx & xxxxxxxxx & xxxxxxxxx 
& xxxxxxxxx & xxxxxxxxx & xxxxxxxxx & xxxxxx & xxxxxx & xxxxxx & xxxxxx & xxxxx & xxxxxxxxxxxxx \cr
{\line{\vrule height0.1pt width19.5cm} 
\vskip 1.0pt
{\line{\vrule height0.1pt width19.5cm }
\+
   Name	& X  &	Y  &  rad & S  & RA (X) & Dec (X) &
 RA (O)& Dec (O)& Dist & B & U-B & Sample & z & COMMENTS \cr
\+
   	& deg  & deg & arcmin  & ergs & H M S & ~$^o$~~ $'$~~ $''$ &  H M S   & ~$^o$~~ $'$~~ $''$  &  arcs.  & mag  & mag  &     &    &     \cr
\+
   	& &  &  & cm$^{-2}$ s$^{-1}$  & & &  & & & & & & &  \cr
\+
(1) & (2) & (3) & (4) & (5) &   (6) &   (7) &   (8) &   (9) & (10) & (11) & (12) & (13) & (14) & (15) \cr
{\line{\vrule height0.1pt width19.5cm }
\settabs \+ xxxxxxxxxxx & xxxxxxx & xxxxxxx & xxxxxxx & xxxxxxxxx & xxxxxxxxx 
& xxxxxxxxx & xxxxxxxxx & xxxxxxxxx & xxxxxx & xxxxxx & xxxxxx & xxxxxx & xxxxx & xxxxxxxxxxxxx \cr
\+
  GSGP4X:24 & $-$0.149 & $-$0.029 &  9.1 & 0.429E-14 & 0 54 23.5 & -27 56 40 &  0 54 23.2 & -27 56 15 &*25.3 & 21.41 &       &   & 0.226 & (W=20A) \cr       
\+
  GSGP4X:48 & $-$0.048 & $+$0.283 & 17.2 & 0.300E-13 & 0 54 51.6 & -27 37 49 &  0 54 51.8 & -27 38  00 & 11.3 & 20.37 & 3.17 & C & 0.155 & (W=37A) \cr          
\+
  GSGP4X:69 & $+$0.029 & $+$0.091 &  5.7 & 0.853E-14 & 0 55 12.2 & -27 49 19 &  0 55 11.5 & -27 49 17 &  9.5 & 20.25 & 3.41 & C & 0.213 & (W=7A, D=1) \cr       
\+
  GSGP4X:72 & $+$0.049 & $+$0.123 &  7.9 & 0.465E-14 & 0 55 17.7 & -27 47 22 &  0 55 20.4 & -27 47 29 & 36.5 & 20.14 & 3.87 &   & 0.316 & (W=5A, D=1) \cr       
\+
  GSGP4X:91 & $+$0.121 & $-$0.242 & 16.2 & 0.350E-13 & 0 55 36.8 & -28 09 15 &  0 55 36.3 & -28 09 23 & 10.4 & 21.33 &      & C & 0.416  & (W=18A) \cr
\+
  GSGP4X:94 & $+$0.138 & $+$0.067  &  9.2 & 0.650E-14 & 0 55 42.0 & -27 50 40 & 0 55 41.9  & -27 50 48 &  8.1 & 20.22 & 2.84 & C & 0.120  & (W=24A) \cr
\+
  GSGP4X:100& $+$0.158 & $-$0.063  & 10.2 & 0.580E-14 & 0 55 47.3 & -27 58 29 & 0 55 48.5  & -27 58 23 & 17.0 & 22.63 & 1.38 &   & 0.597  & (W=51A)  \cr
\+
   SGP3X:26 & $-$0.036 & $+$0.098  & 6.3  & 0.969E-14 & 0 52 25.1 & -28 30 16 & 0 52 27.4  & -28 30 28 & 32.6 & 19.76 & 0.16 &   & 0.162  & (W=42A)  \cr
\+
   SGP3X:30 & $-$0.012 & $+$0.062  &  3.8 & 0.666E-14 & 0 52 31.6 & -28 32 27 & 0 52 31.8  & -28 32 01 & 26.1 & 20.87 &      &   & 0.077  & (W=23A)   \cr
\+
   QSF1X:33 & $-$0.069 & $-$0.100  &  7.3 & 0.566E-14 & 3 40 08.5 & -45 10 17 & 3 40 10.3  & -45 10 20 &*19.3 & 19.57 & 0.23 & C & 0.312  & (W=60A, QSF1:10) \cr
\+
   QSF1X:36 & $-$0.066 & $+$0.264  & 16.3 & 0.227E-13 & 3 40 08.3 & -44 48 14 & 3 40 07.9  & -44 48 24 & 10.9 & 21.07 & 0.17 & C & 0.551  & (W=2A)        \cr
\+
   QSF1X:64 & $+$0.161 & $-$0.012  &  9.7 & 0.852E-14 & 3 41 26.7 & -45 04 44 & 3 41 27.6  & -45 04 31 &*16.1 & 20.20 & 0.61 & C & 0.253  & (W=11A)      \cr
}
}}}
{\line{\vrule height0.1pt width19.5cm} 
\tx\rm

\vfill\eject

\null
.
\vfill\eject




\font\sevenrm=cmr7    \font\fiverm=cmr5
\magnification=\magstep0
\newdimen\digitwidth
\setbox0=\hbox{\rm0}
\digitwidth=\wd0
\catcode`?\active       
\def?{\kern\digitwidth}
\hsize=19.5 true cm
\hoffset=-2.0 true cm
\voffset=0.0 true cm
\vsize=22 true cm
\raggedbottom
\baselineskip 10pt

\table{2}{D}
\noindent
\null
{\centerline {\bf Table 2. \rm ``Ambiguous" EMSS sources reported in Table 8 
and 10 of Stocke et al., 1991}}
\smallskip
\sevenrm
Column 1 gives the X-ray source name,
Column 2 the redshift,
Column 3 the optical V magnitude,
Column 4 the (0.3 - 3.5 keV) X-ray flux in units of $10^{-13}$ \ergcms, 
Column 5 the 5 GHz radio flux in mJy,
Column 6 the classification in the EMSS,
Columns 7, 8, 9, 10 the ratio of selected emission lines where available,
Column 11 the classification based on improved spectroscopy,
Column 12 the reference for the spectroscopic data,
Column 13 notes with a ``C'' the objects included in the statistical analysis.
The sources noted with an ``E'' are ambiguous EMSS sources without
good signal-to-noise spectra;
Column 14 lists the relevant Table number from Stocke et al. (1991).
The abbreviations in column 11 are as follows:
HII   : Starburst-like galaxy;
BLAGN : Broad Line AGN;
Sy2   : Seyfert 2 galaxy.
The references abbreviated in column 12 are as follows:
FG91 : Fruscione and Griffiths (1991); 
R82  : Reichert et al. (1982);
HB92 : Huchra and Burg (1992);
B95b  : Boyle et al. (1995b).
Note that the EMSS source MS1532.5+0130, classified as an AGN by Stocke
et al. (1991), has since been re-classified as a cluster of galaxies
(Maccacaro et al. 1994).  In column 3 ``+'' indicates those objects for
which we have assumed B--V = 1.1 and ``*'' indicates those objects for
which we have assumed B--V = 0.8. For the other objects, we have taken
B--V = 0.3 (see Stocke et al., 1991 for details).

{
\settabs \+ xxxxxxxxxxxxxxx& xxxxxxx & xxxxxx & xxxxxx & xxxxx & xxxxxxxx & xxxxxxxxxx & xxxxxxxxxx & xxxxxxxxxx & xxxxxxxxxx & xxxxxxxxx & xxxxxxxxxxxxx & xxxxxx & xxxxx \cr
{\line{\vrule height0.1pt width19.5cm} 
\vskip 1.0pt
{\line{\vrule height0.1pt width19.5cm }
\+
   Name	& z & $m_V$ & $f_X$ & $f_r$ & ID & log($[OIII]\over H_{\beta}$) & log($[NII]\over H_{\alpha}$) & log($[SII]\over H_{\alpha}$) & log($[OI]\over H_{\alpha}$) & Class. &  Ref & Sample & Comments \cr
\+
        &   &       &       &       & EMSS &                            &                              &                            &                              & \cr                     
\+
(1) & (2) & (3) & (4) & (5) &   (6) &   (7) &   (8) &   (9) & (10) & (11) & (12) & (13) & (14) \cr
{\line{\vrule height0.1pt width19.5cm }
\settabs \+ xxxxxxxxxxxxxxx& xxxxxxx & xxxxxx & xxxxxx & xxxxx & xxxxxxxx & xxxxxxxxxx & xxxxxxxxxx & xxxxxxxxxx & xxxxxxxxxx & xxxxxxxxx & xxxxxxxxxxxxx & xxxxxx & xxxxx \cr
\+
 MS0834.0+6517 & 0.019 & 14.16+& 1.69 & 13.3 & GAL & 0.26 & -0.56 & -0.44 & -1.54 & HII   & FG91       & C & Tab.8  \cr
\+
 MS0942.8+0950 & 0.013 & 15.60 & 42.31& $<$0.6 & AGN &      &       &       &       & BLAGN & This paper &  & Tab.8  \cr                                             \+
 MS1019.0+4836 & 0.062 & 16.93 & 2.70 & $<$0.6 & AGN &      &       &       &       & BLAGN & This paper &  & Tab.8  \cr                                                    \+
 MS1043.9+1400 & 0.010 & 14.53+& 3.93 & 19.5 & GAL & -0.61& -0.13 & -0.47 & -1.16 & HII   & This paper & C & Tab.8  \cr
\+
 MS1047.3+3518 & 0.040 & 15.57 & 9.21 &  2.6 & AGN &  0.38& -0.23 & -0.55 & -1.40 & Sy2   & This paper & C & Tab.8  \cr
\+
 MS1058.8+1003 & 0.028 & 15.47*& 4.62 &  5.9 & AGN & -0.13& -0.22 & -0.46 & -0.98 & HII   & This paper & C & Tab.8  \cr
\+
 MS1110.3+2210 & 0.030 & 16.59 & 5.23 &  1.3 & AGN &      &       &       &       & BLAGN & This paper &  & Tab.8  \cr                                                    \+
 MS1114.4+1801 & 0.092 & 16.50*& 1.02 & $<$0.6 & AGN &      &       &       &       &       &            & E & Tab.8  \cr 
\+
 MS1143.6+2040 & 0.023 & 14.00+& 5.95 &  8.2 & GAL &      &       &       &       & LINER   & R82        & C & Tab.8 \cr
\+
 MS1208.2+3945 & 0.022 & 13.00+& 3.69 & $<$0.5 & GAL &      &       &       &       & LINER & HB92       & C & Tab.8 \cr
\+
 MS2348.6+1956 & 0.043 & 17.12 & 6.31 &  2.2 & AGN & 0.44 & -0.53 & -0.62 & -1.42 & HII   & FG91       & C & Tab.8 \cr
\+
 MS0038.0+3242 & 0.197 & 18.06*& 2.02 & $<$1.1 & AGN &      &       &       &       & BLAGN  & This paper &  & Tab.10 \cr                                                  \+
 MS0038.7+3251 & 0.225 & 18.52*& 3.12 & $<$2.4 & AGN & -0.19& -0.45 &       &       & BLAGN   & This paper &   & Tab.10 \cr
\+
 MS0039.0$-$0145 & 0.110 & 17.86*& 2.30 & $<$0.8 & AGN &      &       &       &       &       &            & E & Tab.10 \cr
\+
 MS0112.9$-$0148 & 0.284 & 18.49*& 1.58 & $<$1.5 & AGN &      &       &       &       &       &            & E & Tab.10 \cr
\+
 MS0116.7+0802 & 0.156 & 17.93 & 1.08 & $<$0.8 & AGN & -0.34& -0.25 & -0.33 & -1.62 & BLAGN    & This paper &   & Tab.10 \cr
\+
 MS0340.3+0455 & 0.097 & 17.76*& 2.06 & $<$0.8 & AGN & 0.45 & -0.22 & -0.13 &       & Sy2 & This paper & C & Tab.10 \cr  
\+
 MS0340.3+0446 & 0.190 & 20.00 & 2.07 & $<$1.0 & AGN &      &       &       &       &       &            & E & Tab.10 \cr 
\+      
 MS0423.8$-$1247 & 0.161 & 19.46*& 1.87 & $<$0.8 & AGN &      & -0.47 & -0.22 & -1.48 & Sy2/HII & This paper & C & Tab.10 \cr                                                   \+
 MS0444.9$-$1000 & 0.095 & 17.50*& 2.50 & $<$0.6 & AGN &      & 0.28 & 0.02 &       & BLAGN & This paper &  & Tab.10 \cr                                                     \+
 MS0850.8+1401 & 0.194 & 17.50*& 9.44 & $<$0.8 & AGN &      &       &       &       &       &            & E & Tab.10 \cr  
\+
 MS0938.1$-$2340 & 0.200 & 19.70 & 1.57 & $<$0.7 & AGN &      &       &       &       &       &            & E & Tab.10 \cr  
\+
 MS1252.4$-$0457 & 0.158 & 19.08 & 2.28 & 4.8  & AGN & 1.046& -0.09 & -0.647 & -1.100 & Sy2   & B95b        & C & Tab.10 \cr
\+
 MS1253.6$-$0539 & 0.420 & 19.97 & 0.94 & $<$5.5 & AGN &      &       &       &       &       &            & E & Tab.10 \cr 
\+
 MS1334.6+0351 & 0.136 & 17.72 & 2.87 & $<$0.7 & AGN & 0.854 & 0.071 & -0.417 & -1.344 & BLAGN  & B95b       &  & Tab.10 \cr
\+
 MS1412.8+1320 & 0.139 & 18.70 & 2.48 & $<$0.6 & AGN & 0.137 & -0.439& -0.476 & -1.545 & HII   & B95b        & C & Tab.10 \cr
\+
 MS1414.8$-$1247 & 0.198 & 19.32 & 7.22 &  6.2 & AGN & 0.699& -0.613& -0.673& -1.367 & HII   & B95b        & C & Tab.10 \cr 
\+
 MS1451.5+2139 & 0.102 & 17.50 & 2.86 & $<$0.8 & AGN &      &       &       &       & Sy2   & S91        & C & Tab.10 \cr
\+
 MS1555.1+4522 & 0.180 & 18.00 & 4.51 & $<$1.0 & AGN & 0.921& -0.013 & -0.253& -0.806  & Sy2   & B95b        & C & Tab.10 \cr 
\+
 MS1614.1+3239 & 0.118 & 17.01*& 2.73 & $<$0.9 & AGN & 1.155 & 0.420& 0.149 & -0.380  & BLAGN   & B95b        &  & Tab.10 \cr 
\+
 MS1804.8$-$6556 & 0.180 & 18.00*& 14.67&      & AGN &      &       &       &       &       &            & E & Tab.10 \cr 
\+
 MS2044.1+7532 & 0.184 & 20.10*& 2.36 & $<$0.8 & AGN & 0.523& -0.104 & -0.314& -1.158  & BLAGN & B95b, This paper&  & Tab.10 \cr 
\+
 MS2118.4$-$1050 & 0.092 & 19.00 & 2.32 & $<$0.7 & AGN &      &       &       &       & BLAGN & This paper   &  & Tab.10 \cr                                             \+
 MS2204.0$-$4059 & 0.231 & 18.60 & 11.49& $<$0.6 & AGN &      &       &       &       &       &            & E & Tab.10  \cr  
\+
 MS2338.9$-$1206 & 0.085 & 18.69 & 2.80 & $<$0.8 & AGN & 0.50 & -0.72 & -0.42 &       & Sy2/HII& This paper & C & Tab.10  \cr        
{\line{\vrule height0.1pt width19.5cm} 
}
}}}
{\line{\vrule height0.1pt width19.5cm} 
\tx\rm

\vfill\eject

\null
.
\vfill\eject




\font\sevenrm=cmr7    \font\fiverm=cmr5
\magnification=\magstep0
\newdimen\digitwidth
\setbox0=\hbox{\rm0}
\digitwidth=\wd0
\catcode`?\active       
\def?{\kern\digitwidth}
\hsize=19.5 true cm
\hoffset=-2.0 true cm
\voffset=0.0 true cm
\vsize=22 true cm
\raggedbottom
\baselineskip 10pt

\table{2}{D}
\noindent
\null
{\centerline {\bf Table 3. \rm Percentage contribution of the NELGs 
to the 2 keV X-ray Background}}
\smallskip

{
\settabs \+ xxxxxxxxxxxxxxx & xxxxxxxxxx & xxxxxxxxxx & xxxxxxxxxx & xxxxxxxxxx & \cr
{\line{\vrule height0.1pt width19.5cm} 
\vskip 1.0pt
{\line{\vrule height0.1pt width19.5cm }
\+
  $log(L_x)$	& $z$ & $z$ & $z$ & $z$ \cr
\+
   	& 0.0--1.0 & 1.0--2.0 & 2.0--3.0 & 0.0--3.0 \cr
{\line{\vrule height0.1pt width19.5cm }
\settabs \+ xxxxxxxxxxxxxxx & xxxxxxxxxx & xxxxxxxxxx & xxxxxxxxxx & xxxxxxxxxx & \cr
\+
    40--41        & 0.29--4.08  & 0.35--4.90 & 0.39--5.51 & 1.03--14.49  \cr
\+  
    41--42        & 0.73--3.24  & 0.88--3.89 & 0.99--4.38 & 2.60--11.51  \cr
\+
    42--43        & 1.67--2.46  & 2.01--2.95 & 2.27--3.33 & 5.95--8.74   \cr
\+
    43--44        & 0.27        & 0.32       & 0.36       & 0.95         \cr
\+ 
    44--47        & $< 0.01$    & $< 0.01$   & $< 0.01$   &  $< 0.01$     \cr
{\line{\vrule height0.1pt width19.5cm }
\+ 
    40--47        & 2.96--10.05 & 3.56--12.06 & 4.01--13.58 & 10.53--35.69 \cr
}
}}}
{\line{\vrule height0.1pt width19.5cm} 
\tx\rm

\vfill\eject

\end